\pdfoutput=1
\documentclass[a4paper,11pt]{article}

\newcommand\papertitle{\boldmath Symmetry-Resolved Entanglement in AdS${}_3$/CFT${}_2$ coupled to  $U(1)$ Chern-Simons Theory}

\usepackage{jheppub}
\usepackage{xcolor}

\usepackage{tikz}
\usetikzlibrary{arrows, backgrounds, calc, decorations.pathreplacing, decorations.markings, fadings, quotes, positioning, shadings, intersections, angles}

\usepackage{amsmath,amsfonts,amssymb,mathrsfs}
\usepackage{bbm}
\usepackage[inline]{enumitem}
\usepackage{multirow}
\usepackage{subcaption}
\usepackage{nicefrac}
\usepackage{xspace}
\usepackage{mathtools}

\usepackage[colorlinks=true]{hyperref}
\hypersetup{
	bookmarks=true,         
	unicode=false,          
	pdftoolbar=true,        
	pdfmenubar=true,        
	pdffitwindow=false,     
	pdfstartview={FitH},    
	pdftitle={\papertitle}, 
	pdfauthor={Christian Northe},
	pdfnewwindow=true,      
	colorlinks=true,        
	linkcolor=blue,         
	citecolor=red,          
	filecolor=magenta,      
	urlcolor=cyan           
}

\makeatletter

\@addtoreset{equation}{section}
\makeatother

\newcommand{\Z}	{\mathbb{Z}}

\newcommand{\cA}{\mathcal{A}}
\newcommand{\cB}{\mathcal{B}}

\newcommand{\cL}{\mathcal{L}}
\newcommand{\cM}{\mathcal{M}}

\newcommand{\cR}{\mathcal{R}}

\newcommand{\cZ}{\mathcal{Z}}

\DeclareMathOperator{\Tr}{Tr}
\DeclareMathOperator{\tr}{\tr}

\newcommand{\normord}[1]{:\mathrel{#1}:}

\renewcommand{\Tr}	{\mathrm{Tr}}
\renewcommand{\tr}	{\mathrm{tr}}

\newcommand\Algebra[1]	{\mathfrak{#1}}

\renewcommand{\u}	{\hat{\Algebra{u}}}

\newcommand\ads {\text{AdS}}
\newcommand\cft {\text{CFT}}

\newcommand{\charge}{\alpha}
\newcommand{\appq}  {\mathsf{q}}
\newcommand{\diffq} {\ell}

\newcommand{\ith}   {i^{\text{th}}}

\newcommand\Secref[1]	{Section~\ref{#1}\xspace}

\newcommand\Appref[1] {Appendix~\ref{#1}\xspace}
\newcommand\secref[1]	{section~\ref{#1}\xspace}

\newcommand\figref[1]	{figure~\ref{#1}\xspace}
\newcommand\appref[1] {appendix~\ref{#1}\xspace}

\begin{document}
	\title{\papertitle}
	
	\author[a,1]{Suting Zhao,\note[1]{suting.zhao@physik.uni-wuerzburg.de}}
	\author[a,2]{Christian Northe,\note[2]{christian.northe@physik.uni-wuerzburg.de}}
	\author[a,3]{Ren\'e Meyer\note[3]{Corresponding author: rene.meyer@physik.uni-wuerzburg.de\\
			The ordering of authors is chosen to reflect their role in the preparation of this work.}}
	
	\affiliation[a]{Institut f\"ur Theoretische Physik und Astrophysik\\ and \\
		W\"urzburg-Dresden Cluster of Excellence ct.qmat,\\ Julius-Maximilians-Universit\"at W\"urzburg, \\Am Hubland \\97074 W\"urzburg, Germany}
	
	\abstract{
		We consider symmetry-resolved entanglement entropy in AdS${}_3$/CFT${}_2$ coupled to $U(1)$ Chern-Simons theory. We identify the holographic dual of the charged moments in the two-dimensional conformal field theory as a charged Wilson line in the bulk of AdS${}_3$, namely the Ryu-Takayanagi geodesic minimally coupled to the $U(1)$ Chern-Simons gauge field. We identify the holonomy around the Wilson line as the Aharonov-Bohm phases which, in the two-dimensional field theory, are generated by charged $U(1)$ vertex operators inserted at the endpoints of the entangling interval. 
		Furthermore, we devise a new method to calculate the symmetry resolved entanglement entropy by relating the generating function for the charged moments to the amount of charge in the entangling subregion. We calculate the subregion charge from the $U(1)$ Chern-Simons gauge field sourced by the bulk Wilson line. We use our method to derive the symmetry-resolved entanglement entropy for Poincar\'e patch and global AdS${}_3$, as well as for the conical defect geometries. In all three cases, the symmetry resolved entanglement entropy is determined by the length of the Ryu-Takayanagi geodesic and the Chern-Simons level $k$, and fulfills equipartition of entanglement. The asymptotic symmetry algebra of the bulk theory is of $\u{(1)_k}$ Kac-Moody type. Employing the $\u{(1)_k}$ Kac-Moody symmetry, we confirm our holographic results by a calculation in the dual conformal field theory.  
	}
	
	
	\maketitle
	\flushbottom

	\section{Introduction}\label{sec:Intro}
	A promising approach to the problem of quantum gravity is via the holographic principle \cite{tHooft,Susskind}. A concrete realization of this principle is provided by the AdS/CFT correspondence \cite{Maldacena}, which originated from string theory. In a particular limit, the AdS/CFT correspondence (often also simply called holography) relates strongly coupled quantum field theories to weakly interacting gravitational theories in asymptotically Anti de Sitter space-times. The Ryu-Takayanagi prescription \cite{RT} for the holographic computation of entanglement entropy led to substantial progress in the study of quantum gravity by establishing a bridge between the AdS/CFT correspondence and quantum information theory. From this, holographers have gained the fundamental idea that the bulk space-time in AdS/CFT should emerge from the  geometrization of core concepts of quantum information such as entanglement entropy \cite{CalabreseCardy}, mutual information \cite{RangamaniTakayanagiBook}, relative entropy \cite{MyersCasiniBlanco}, quantum computational complexity \cite{Susskind:2014rva,Brown:2015bva,Brown:2015lvg}, tensor networks \cite{Swingle:2009bg,Happy},  and quantum error correction \cite{HarlowAlmheiriDong}. Moreover, the quantum information community has profited as well from this connection, one instance being the refinement of our understanding of operator algebra quantum error correction \cite{PreskillPastawski}.
	
	Recently, a more refined notion of entanglement entropy was introduced in \cite{GoldsteinSela}, applicable whenever additional conserved charges exist in a system. The question approached in \cite{GoldsteinSela} was how an Aharanov-Bohm flux $\alpha$ inserted on the replica manifold influences a charged particle. It was noted that the particle acrues a phase proportional to the total charge in subregion $A$. In formalizing this observation, the authors of \cite{GoldsteinSela} were led to introduce a more refined notion of entanglement,  the \textit{symmetry resolved entanglement entropy}. Given a global symmetry, the Hilbert space organizes into representations of the symmetry, or charge sectors. The symmetry-resolved entanglement entropy quantifies the entanglement associated with the sector of fixed subregion charge.
	
	The aim of this paper is two-fold. Our first goal is to establish a new method for the computation of the symmetry-resolved entanglement entropy, complementary to the approach pioneered in \cite{GoldsteinSela}. We show that the symmetry-resolved entanglement entropy can be deduced from a generating function. This approach has the virtue of allowing the determination of the charged contributions of the symmetry-resolved entanglement entropy directly from the expectation value of the subregion charge operator $Q_\cA$. Another advantage over the previous approaches \cite{GoldsteinSela,Bonsignori:2019naz,Murciano:2020lqq,Capizzi:2020jed,Murciano:2020vgh} lies in the study of excited states, where our method bypasses the potentially laborious computation of charged moments\footnote{In the context of symmetry resolved R\'enyi entropies, the charged moments are the analogues of the grand canonical partition function in statistical physics, in which the chemical potential is fixed and the charge is allowed to fluctuate. In the charged moments, it is the chemical potential dual to the subregion charge inside the entangling region that is kept fixed when calculating the symmetry resolved R\'enyi entropies.} altogether. 
	
	This new formalism is used to achieve our second main goal in this paper: We provide a new geometric realization of the symmetry-resolved entanglement entropy within gauge/gravity duality. Our setup is $\ads_3/\cft_2$ holography coupled to  $U(1)$ Chern-Simons theory. The $U(1)$ Chern-Simons gauge fields are dual to a conserved $U(1)$ current, and hence a conserved $U(1)$ charge, in the boundary conformal field theory. This model is interesting, because it is non-trivial yet integrable. It carries a $\u(1)_k$ Kac-Moody algebra, turning the charged sectors into familiar affine $U(1)$ representations, which are well-understood on the gravity as well as the field theory side. Moreover, the Chern-Simons gauge fields  decouple from gravity, such that exact solutions for the gauge fields can be found. Our geometric realization of the symmetry resolved entanglement involves a $U(1)$ Wilson line defect in the bulk, anchored at the endpoints of the entangling region in the boundary. In the replica manifold, this Wilson line defect  follows  the fixed point locus of the replica symmetry, and hence is identified as the Ryu-Takayanagi geodesic minimally coupled to the $U(1)$ Chern-Simons gauge field. At the endpoints of the entangling region, the Wilson line generates an insertion of Aharonov-Bohm flux \cite{GoldsteinSela}. Calculating the subregion charge on the boundary from the bulk Chern-Simons gauge fields sourced by the Wilson line defect, our generating function method allows us to determine the symmetry resolved entanglement entropy. We apply our holographic construction to general asymptotically $\ads_3$ geometries \cite{Banados:1994tn}. We then cover the examples of Poincar\'e and global $\ads_3$, as well as conical defects. In all these cases, the symmetry resolved entanglement entropy is equally distributed in different charge sectors, which is known as the equipartition of the entanglement entropy \cite{Xavier:2018kqb}.

	All CFT calculations \cite{GoldsteinSela,Bonsignori:2019naz,Murciano:2020lqq,Capizzi:2020jed,Murciano:2020vgh} of symmetry resolved entanglement so far use the charged moments in their computation. Also in the holographic setup, the charged moments  have been discussed in \cite{Belin:2013uta}, where, for the vacuum state, they were related to a topological black hole by a  conformal transformation. The thermal state described by the topological black hole has a natural $U(1)$ symmetry, the time evolution around the Euclidean time circle. The charged moments are then simply the partition function of the topological black hole. For excited states, this approach is rendered invalid by the additional insertions in the boundary CFT, which break the $U(1)$ symmetry generated by the modular Hamiltonian. In more general holographic setups including for example excited states, our generating function approach, which does not rely on the charged moments, may be more applicable.
	
	We confirm our gravity results with an independent calculation in two-dimensional conformal field theory, which provides a further non-trivial test of the $\ads/\cft$ correspondence. The bulk Wilson line is dual to a charge defect in the $\cft$, which is naturally described by a pair of $U(1)$ vertex operators  inserted at the endpoints of the entangling region $\cA$. The vertex operators are charged under the $\u(1)_k$ Kac-Moody symmetry of the dual conformal field theory. Similar to the conventional replica trick involving twist fields inserted at the endpoints of the entangling region, the $U(1)$ vertex operators are connected by a branch cut along $\cA$, which can be thought of as a charge defect line. It induces monodromies on charged fields upon crossing the defect line. These monodromies are the desired Aharanov-Bohm phases, and are related to the holonomies of the $U(1)$ Chern-Simons field around the bulk Wilson line. Similar relations have been explored in the first order Chern-Simons formulations of gravity \cite{deBoer:2014sna}, where the monodromies induced by the twist operators on the boundary are directly related to the holonomies of the Chern-Simons connections in the bulk. Using the charged defect, we directly compute the symmetry-resolved entanglement entropy  in the $\cft$ vacuum state, and find perfect agreement with our holographic result.
	
	Our paper is structured as follows. In \Secref{section 2}, we first review the formalism of \cite{GoldsteinSela}, and then present our generating function approach. In \secref{section 3} we introduce $\ads_3$ gravity coupled to $U(1)$ Chern-Simons gauge fields. We relate the charged moments necessary for the computation of the symmetry-resolved entanglement entropy to the $U(1)$ Wilson lines in $\ads_3$. We present our examples in \secref{section 4}, in particular Poincar\'e $\ads$, global $\ads$, and the conical defect geometries. In \secref{section 5} we confirm our holographic results within conformal field theory. We conclude and give an outlook in \secref{Section 6}. Details on the asymptotic symmetry algebra and on vertex operators are relegated to the appendices.
	
	\section{Symmetry resolved entanglement}\label{section 2}
	Given a system with a global symmetry $G$, the spectrum decomposes into corresponding representations. Each representation corresponds to a charged sector.  In this situation it is pssible to investigate the entanglement associated with each of these charge sectors, the symmetry resolved entanglement entropy. \Secref{section 2.1} provides a short review of general aspects of the symmetry-resolved entanglement entropy. In \secref{section 2.2} we develop a new method to calculate the symmetry-resolved entanglement entropy based on a generating function. 
	\subsection{Entanglement in charge sectors}\label{section 2.1}
	We consider a system with an internal $U(1)$ symmetry and its bipartition into two subsystems, $\mathcal{A}$ and its complement $\mathcal{B}$. The charge operator $Q$ is the generator of the symmetry and we assume that $Q=Q_{\mathcal{A}}\oplus Q_{\mathcal{B}}$. If the system is in an eigenstate of $Q$,%
	\footnote{This is not an assumption, since a global conserved charge in quantum theory leads to the decomposition of the Hilbert space into superselection sectors labelled by charge eigenvalues \cite{Giulini:2007fn}. This in particular implies a restriction on the superposition principle of quantum mechanics, forbidding superposition of states with distinct conserved charges.}
	 then $[\rho,Q]=0$. Tracing out the degrees of freedom of $\cB$, one obtains $[\rho_{\mathcal{A}},Q_{\mathcal{A}}]=0$. Hence $\rho_{\mathcal{A}}$ has a block-diagonal structure. Each block of $\rho_{\mathcal{A}}$ corresponds to an eigenvalue $q$ of the subregion charge operator $Q_{\mathcal{A}}$. 
	\begin{eqnarray}
		\rho_{\mathcal{A}}=\oplus_{q}\rho_{\mathcal{A}}(q)\ ,\label{sum of rhoAq}
	\end{eqnarray}
	The block $\rho_\cA(q)$ is selected by the projector $\Pi_q$ onto the eigenspace of $Q_\cA$ with fixed eigenvalue $q$,
	\begin{eqnarray}
		\rho_{\mathcal{A}}(q)=\rho_{\mathcal{A}}\Pi_{q}\ .
	\end{eqnarray}	
	The probability of finding eigenvalue $q$ in a measurement of $Q_{\mathcal{A}}$ is given by
	\begin{eqnarray}
		P_{\mathcal{A}}(q)=\frac{\Tr\rho_{\mathcal{A}}(q)}{\Tr\rho_{\mathcal{A}}}=\Tr\rho_{\mathcal{A}}(q)\ .
	\end{eqnarray}
	since $\rho_\cA$ is normalized, $\Tr \rho_{\mathcal{A}}=1$. Furthermore, the block-diagonal structure of $\rho_{\mathcal{A}}$ induces block decomposition on $\rho_{\mathcal{A}}^n$,
	\begin{eqnarray}\label{sum of rhoAqn}
		\rho_{\mathcal{A}}^{n}=\oplus_{q}\rho_{\mathcal{A}}(q)^n
		\quad
		\text{with}
		\quad
		\rho_{\mathcal{A}}(q)^n=\left(\rho_{\mathcal{A}} \Pi_{q}\right)^n=\rho_{\mathcal{A}}^n\Pi_q\, .
	\end{eqnarray}
	The probability distributions of different charge blocks in $\rho_{\mathcal{A}}^n$ are given by
	\begin{eqnarray}
		P_{\mathcal{A},n}(q)=\frac{\Tr\rho_{\mathcal{A}}(q)^n}{\Tr\rho_{\mathcal{A}}^n}\ ,\label{Pn}
	\end{eqnarray}
	Given the decompositions \eqref{sum of rhoAq} and \eqref{sum of rhoAqn}, one defines the \textit{symmetry-resolved entanglement entropy} and \textit{symmetry-resolved R\'enyi entropy}, which are a measure of the amount of entanglement between the subsystems $\mathcal{A}$ and $\mathcal{B}$ in each of these blocks. The symmetry-resolved R\'enyi entropy is defined as
	\begin{eqnarray}
		S_{n}(q)=\frac{1}{1-n}\log \Tr\left(\frac{\rho_{\mathcal{A}}(q)}{P_{\mathcal{A}}(q)}\right)^{n}=\frac{1}{1-n}\log{\frac{\mathcal{Z}_{n}(q)}{\mathcal{Z}_{1}(q)^{n}}}\ ,
	\end{eqnarray}
	where we denote
	\begin{eqnarray}
		\mathcal{Z}_{n}(q)=\Tr\rho_{\mathcal{A}}(q)^n=\Tr \left[\rho_{\mathcal{A}}^n\Pi_q\right]\ ,\label{Znq}
	\end{eqnarray}
	representing the contribution to the partition function $\mathcal{Z}_n$ from the $q$-block. The symmetry resolved entanglement entropy in the $q$-block can be obtained by taking the $n\to 1$ limit of $S_{n}(q)$, which reads
	\begin{eqnarray}
		S_{1}(q)=\lim_{n\to 1} S_{n}(q)=-\Tr\left[ \frac{\rho_{\mathcal{A}}(q)}{P_{\mathcal{A}}(q)}\log\left(\frac{\rho_{\mathcal{A}}(q)}{P_{\mathcal{A}}(q)}\right)\right]\ . \label{S1q}
	\end{eqnarray}
	From \eqref{sum of rhoAq} and \eqref{S1q}, the total entanglement entropy associated with $\rho_{\mathcal{A}}$ is given by
	\begin{eqnarray}
		S_{1}=\sum_{q}P_{\mathcal{A}}(q)S_{1}(q)-\sum_{q}P_{\mathcal{A}}(q)\log P_{\mathcal{A}}(q)\ ,\label{S1-S1q}
	\end{eqnarray}
	where the two terms in \eqref{S1-S1q} are usually referred to as \textit{configurational} and \textit{fluctuation} entropy, respectively \cite{Lukin256}. In particular, the configurational entropy is also related to the operationally accessible entanglement entropy \cite{WisemanVaccaro, BarghathiHerdman, BarghathiCasiano}.
	
	To obtain the symmetry-resolved R\'enyi entropy and entanglement entropy, one needs to calculate $\mathcal{Z}_{n}(q)$. However, this is generally very difficult in general, since it requires the knowledge of the spectrum of the reduced density matrix $\rho_{\mathcal{A}}$ and its resolution in $Q_{\mathcal{A}}$.
	Rather than calculating $\cZ_n(q)$ directly, the idea advocated in \cite{GoldsteinSela} is to focus on the computation of the \textit{charged moments}
	\begin{eqnarray}\label{chargedMoments}
		\mathcal{Z}_{n}(\mu)=\Tr\left[\rho_{\mathcal{A}}^{n}e^{i\mu Q_{\mathcal{A}}}\right]
	\end{eqnarray}
	From \eqref{Znq}, $\mathcal{Z}_{n}(q)$ can be obtained by a Fourier transformation of the charged moments,
	\begin{eqnarray}\label{chargedMomentsFourier}
		\mathcal{Z}_{n}(q)=\int_{-\infty}^{\infty}\frac{d\mu}{2\pi} e^{-i\mu q}\mathcal{Z}_{n}(\mu)
	\end{eqnarray}
	Here we have assumed that the eigenvalues of $Q_{\mathcal{A}}$ are continuous. In the case that the eigenvalues are integers, i.e. $\mathcal{Z}_{n}(\mu)=\mathcal{Z}_{n}(\mu+2\pi)$, one needs to change the range of integration in \eqref{chargedMomentsFourier} to $[-\pi,\pi]$.
	
	\subsection{Generating function method}\label{section 2.2}
	Even though the charged moments \eqref{chargedMoments} provide a way to calculate the symmetry-resolved entropy, it's still difficult to compute $\mathcal{Z}_{n}(\mu)$ for general excited states (c.f. \cite{Capizzi:2020jed, Horvath:2020vzs, Bonsignori:2020laa}). In particular, the holographic calculation requires constructing the holographic dual of the CFT charged moments, and evaluating the on-shell action of this configuration. We introduce here a method based on a generating function, which simplifies the calculation of the symmetry-resolved entanglement entropy, as it reveals how to compute the symmetry-resolved entanglement entropy directly from the expectation value of the subregion charge operator $Q_\cA$. The latter is generically easier to compute than the charged moments. Examples are presented in subsequent sections in the context of holography.
	
	We start by defining the following normalized generating function associated with the charged moments $\mathcal{Z}_{n}(\mu)$ as\footnote{A similar but not completely equivalent ratio of partition functions has been employed in \cite{Bonsignori:2019naz,Capizzi:2020jed}. Also, the calculation in \cite{Bonsignori:2019naz,Capizzi:2020jed} was done by directly evaluation the charged moments in the theories considered there. Our method evades the evaluation of the charged moments altogether.}
	\begin{eqnarray}\label{generatingfunction}
		f_{n}(\mu)\coloneqq \frac{\mathcal{Z}_{n}(\mu)}{\mathcal{Z}_{n}(0)}=\frac{\mathcal{Z}_{n}(\mu)}{\mathcal{Z}_{n}},
	\end{eqnarray}
	with initial condition 
	\begin{equation}\label{initialCondition}
		f_{n}(0)=1\,.
	\end{equation}
	The expectation value of $Q_A$ can then be expressed as
	\begin{eqnarray}\label{chargeExpectationValue}
		\langle iQ_{\mathcal{A}}\rangle_{n,\mu}
		:=
		\frac{\Tr \left[iQ_{\mathcal{A}} \rho^{n}e^{i\mu Q_{\mathcal{A}}}\right]}{\Tr\left[\rho^{n}e^{i\mu Q_{\mathcal{A}}}\right]}
		=
		\frac{\partial \ln \cZ_{n}(\mu)}{\partial\mu}
		=
		\frac{\partial \ln f_{n}(\mu)}{\partial \mu}\ .\label{Q-f}
	\end{eqnarray}
	In general, $Z_n(\mu)$ is already known from the usual calculations of the R\'enyi entropies. In order to calculate $f_n(\mu)$, and hence $Z_n(\mu)$, it suffices to know the expectation value of the charge, and integrate \eqref{chargeExpectationValue} with initial condition \eqref{initialCondition}.
	
	The symmetry-resolved R\'enyi entropy can be rewritten in terms of the generating function $f_n(\mu)$. As evident from \eqref{Pn}, \eqref{Znq} and  \eqref{chargedMomentsFourier}, the probability distribution $P_{\mathcal{A},n}(q)$ is related to the generating function by a Fourier transformation,
	\begin{eqnarray}\label{probabilityGeneratingFunction}
		P_{\mathcal{A},n}(q)=\frac{\mathcal{Z}_n(q)}{\mathcal{Z}_n}=\int_{-\infty}^{\infty}\frac{d\mu}{2\pi} e^{-i\mu q}f_{n}(\mu)\ ,
	\end{eqnarray}
	As pointed out in \cite{Capizzi:2020jed}, the symmetry-resolved R\'enyi entropy,
	\begin{eqnarray}
		S_{n}(q)=S_n+\frac{1}{1-n}\log{\frac{P_{\mathcal{A},n}(q)}{P_{\mathcal{A}}(q)^{n}}}\ ,\label{SRE simp}
	\end{eqnarray}
	as well as the symmetry-resolved entanglement entropy
	\begin{eqnarray}
		S(q)=S+\lim_{n\to 1}\frac{1}{1-n}\log{\frac{P_{\mathcal{A},n}(q)}{P_{\mathcal{A}}(q)^{n}}}\ ,\label{SEE simp}
	\end{eqnarray}
	decompose into two additive contributions. The first are their uncharged counterparts, the R\'enyi entropy and the entanglement entropy, while the charge information resides fully in the second contribution. Remarkably, our generating function approach traces the computation of the latter all the way back to $\langle Q_\cA\rangle$, c.f. \eqref{chargeExpectationValue} and \eqref{probabilityGeneratingFunction}. Furthermore, this method has practical merits, as for general excited states the expectation value of $Q_{\mathcal{A}}$ is much easier to calculate than $\mathcal{Z}_{n}(\mu)$.
	
	Our approach is particularly efficient in holography, since we already know that the entanglement entropy is given by the Ryu-Takayanagi formula \cite{RT}. The remaining task in the holographic calculation is now reduced from finding the exact holographic dual of the CFT charged moments $\mathcal{Z}_n(\mu)$, to simply obtaining the boundary expectation value of $Q_{\mathcal{A}}$. As we show in Section \ref{section 3.2} this holds true even without knowing the exact holographic renormalization, as long as the proposed bulk state contains the same amount of subregion charge $Q_{\mathcal{A}}$ as the exact dual. In this case, the information leading to the symmetry-resolved entanglement entropy can still be  extracted from the proposed bulk state.
	
	\section{Holographic U(1) Chern-Simons-Einstein gravity}\label{section 3}
	
	In this section, we begin by briefly reviewing some aspects of $AdS_3$ gravity with $U(1)$ Chern-Simons gauge fields relevant for our work. This model, under appropriate boundary conditions, contains two chiral $U(1)$ Kac-Moody currents on the boundary, which provide the notion of the $U(1)$ charge. Our main task is then to investigate the gravity dual of the charged moments. The basic idea is that the operator $e^{i\mu \hat{Q}_{\cA}}$ in the charged moments is generated by additional source terms on the gravity side, which is coupled to the Kac-Moody currents. For this reason, we introduce a $U(1)$ Wilson line defect in the bulk. After solving the gauge fields with defect insertion, we calculate the shift of the deformation to the CFT action stemming from the defect, and show that indeed this holographic construction realizes the gravity dual of the charged moments in general asymptotic $\ads$ backgrounds. In particular, these allow us to obtain the subregion charge of the charged moments from the boundary values of the gauge fields. We defer the derivation of the symmetry-resolved entanglement entropy to \secref{section 4}.
	
	\subsection{$\ads_3$ gravity with U(1) Chern-Simons fields}\label{section 3.1}
	Considering the three-dimensional Einstein-gravity with negative cosmological constant as well as additional $U(1)$ Chern-Simons terms on a manifold $\mathcal{M}$. The topology of the manifold is $\Sigma\times \mathbbm{R}$, where $\Sigma$ denotes the constant Euclidean time slice. The total Euclidean action is given by
	\begin{eqnarray}
		I_{0}= I_g+I_A+I_{\bar{A}}\ .
	\end{eqnarray}
	The Euclidean gravity action $I_g$ contains the Einstein-Hilbert action, the boundary Gibbons-Hawking term, and the volume counter term \cite{Balasubramanian:1999re},
	\begin{eqnarray}
		I_g=-\frac{1}{16\pi G_3}\int_{M} dx^3 \sqrt{g} \left(R-\frac{2}{l^{2}}\right)-\frac{1}{8\pi G_3}\int_{\partial M}dx^2 \sqrt{h} (K-\frac{1}{l})\ ,
	\end{eqnarray}
	with $l$ being the AdS radius. For later convenience, we set $l=1$. The action for the gauge fields contain two chiral sectors, given by the usual $U(1)$ Chern-Simons terms as well as additional boundary terms \cite{Kraus:2006nb, Kraus:2006wn},
	\begin{eqnarray}\label{CSaction}
		I_A&=&\frac{i k}{8\pi}\int_{\mathcal{M}} A\wedge d A-\frac{k}{16\pi}\int_{\mathcal{\partial M}} dx^2 \sqrt{g}A^{i}A_{i}\ ,\nonumber\\
		I_{\bar{A}}&=&-\frac{i k}{8\pi}\int_{\mathcal{M}} {\bar{A}}\wedge d {\bar{A}}-\frac{k}{16\pi}\int_{\mathcal{\partial M}} dx^2 \sqrt{g}{\bar{A}}^{i}{\bar{A}}_{i}\ ,\label{A action}
	\end{eqnarray}
	where the dimensionless Chern-Simons level $k$ can take any real value.\footnote[1]{In the case that the gauge group $G$ is non-abliean and compact, the level $k$ is quantized to be integer\cite{Dunne:1998qy}.} In Fefferman-Graham coordinates, the metric solution takes the following asymptotic form as $\rho\to \infty$,
	\begin{eqnarray}
		ds^2=d\rho^2+e^{2\rho}\left(g_{ij}^{(0)}+e^{-2\rho}g_{ij}^{(2)}+\cdots\right)dx^i dx^j, \quad i, j=1, 2\ .
	\end{eqnarray}
	The variation of the gravity action with respect to $g^{(0)}_{ij}$,
	\begin{eqnarray}
		\delta I_g=\frac{1}{2}\int dx^2 \ T^{ij}[g]\delta g^{(0)}_{ij}\ ,
	\end{eqnarray}
	yields the renormalized gravitational stress tensor \cite{Balasubramanian:1999re}
	\begin{eqnarray}\label{g tensor}
		T_{ij}[g]=\frac{1}{8\pi G_3 l}\left(g^{(2)}_{ij}-\Tr g^{(2)}g^{(0)}_{ij}\right)\ .
	\end{eqnarray}
	In the pure gravity case, $T_{ij}[g]$ identified as the expectation value of the stress tensor in dual CFT, and $g^{(0)}_{ij}$ represents the boundary metric.
	
	To present the asymptotic symmetry algebrae, in this subsection, we consider a conformal boundary with topology $S^1\times \mathbbm{R}^1$. The boundary complex coordinates are defined as $w=\phi+i \tau$ and $\bar{w}=\phi-i\tau$, with $\tau$ as the Euclidean time, and $\phi$ as the compact spatial direction, i.e. $\phi\sim\phi+2\pi$.\footnote[2]{In later sections, we will also consider the Poincar\'e patch. In order to distinguish two different cases, the complex coordinates in the Poincar\'e patch are defined as $z=x+i t_E$ and $\bar{z}=x-i t_E$, and the definitions for the current modes should be changed accordingly.} Various types of boundary conditions for $\ads_3$ gravity lead to different asymptotic symmetry algebrae \cite{Brown:1986nw, Compere:2013bya, Donnay:2015abr, Afshar:2015wjm, Afshar:2016wfy, Grumiller:2016pqb,Melnikov:2018fhb}. Here we choose the Brown-Henneaux boundary condition for the metric\cite{Brown:1986nw}, 
	\begin{eqnarray}
		g^{(0)}_{ij}dx^i dx^j= dw d\bar{w}\ .\label{BH condition}
	\end{eqnarray}
	In this case, the stress tensor $T_{ww}[g]$ $(T_{\bar{w}\bar{w}}[g])$ is then restricted to an arbitrary holomorphic (anti-holomorphic) function and the asymptotic symmetry algebrae are given by two copies of the Virasoro algebra. To be precise, the Brown-Henneaux boundary condition \eqref{BH condition} is preserved by the following infinitesimal diffeomorphisms,
	\begin{eqnarray}
		w&\to& w+\xi(w)-\frac{1}{2}e^{-2\rho}\partial_{\bar{w}}^2 \bar{\xi}(\bar{w})\ ,\nonumber\\
		\bar{w}&\to& \bar{w}+\bar{\xi}(\bar{w})-\frac{1}{2}e^{-2\rho}\partial_{w}^2 \xi(w)\ ,\nonumber\\
		\rho&\to&\rho-\frac{1}{2}\left(\partial_w\xi(w)+\partial_{\bar{w}}\bar{\xi}(\bar{w})\right)\ .\label{asy diffeo}
	\end{eqnarray}
	where $\xi(w)$ and $\bar{\xi}(\bar{w})$ are arbitrary functions. The diffeomorphisms \eqref{asy diffeo} act non-trivially on $g^{(2)}_{ij}$, and transform the renormalized gravitational stress tensor \eqref{g tensor} as
	\begin{eqnarray}
		T_{ww}[g] \to T_{ww}[g]+2\partial_w\xi(w)T_{ww}[g]+\xi(w)\partial_w T_{ww}[g]-\frac{c}{24\pi}\partial_{w}^3\xi(w)\ .
	\end{eqnarray}
	This is exactly the conformal transformation law for a CFT stress tensor with the Brown-Henneaux central charge $c=3l/2G_3$. As shown in \cite{Brown:1986nw}, using the Brown-Henneaux boundary conditions, the modes of the stress tensor, defined as
	\begin{eqnarray}
		\Tilde{L}_n-\frac{c}{24}\delta_{n,0}&=&-\oint dw e^{-inw}T_{ww}[g],\nonumber\\
		\Tilde{\bar{L}}_n-\frac{c}{24}\delta_{n,0}&=&-\oint dw \ e^{-in\bar{w}} \bar{T}_{\bar{w}\bar{w}}[g],
	\end{eqnarray}
	yield the Virasora algebra
	\begin{eqnarray}
		[\Tilde{L}_n,\Tilde{L}_m]=(n-m)\Tilde{L}_{n+m}+\frac{c}{12}(n^3-n)\delta_{m+n,0}\ .\label{cl vir}
	\end{eqnarray}
	
	For the $U(1)$ gauge fields, we focus on the left-mover $A$. The analysis follows through for the right mover $\bar{A}$. In Fefferman-Graham coordinates, the solutions $A$ are supposed to take the following form \cite{Kraus:2006wn}
	\begin{eqnarray}
		A=A^{(0)}+e^{-2\rho}A^{(2)}+\cdots\ , \quad \rho\to\infty\ .\label{A expansion}
	\end{eqnarray}
	Since the topology in the radial direction is trivial, one can always perform a gauge transformation such that the leading order of $A^{(0)}_{\rho}$ vanishes. Therefore we impose the boundary condition $A^{(0)}_{\rho}=0$, which enforces $A^{(0)}_i$ to be a flat connection, as required by the equation of motion. Varying the action \eqref{CSaction} with respect to $A$ yields \cite{Kraus:2006wn}
	\begin{eqnarray}
		\delta I_{A}&=&\frac{i}{2\pi}\int_{\mathcal{\partial M}} d x^2 \sqrt{g^{(0)}} J^{a}\delta A_{a}^{(0)}\ .\label{var A}
	\end{eqnarray}
	In complex coordinates, the components of the current $J$ read
	\begin{eqnarray}
		J_{w}=\frac{1}{2}J^{\bar{w}}=\frac{i k}{2}A^{(0)}_{w}\ ,\quad J_{\bar{w}}=0\ .\label{current component}
	\end{eqnarray}
	Following from \eqref{var A}, the well-defined variational principle is to allow $A^{(0)}_w$ to vary while keeping $A^{(0)}_{\bar{w}}$ fixed. Furthermore, as we show in \appref{appendix A}, the modes of the $U(1)$ current, defined as
	\begin{eqnarray}
		J_{n}=\oint\frac{d w}{2\pi } e^{-inw} J_{w}\ ,\label{current modes}
	\end{eqnarray}
	fulfill an affine $\u(1)_{k}$ Kac-Moody algebra \cite{Kraus:2006nb}.
	
	Since the bulk and boundary symmetries match by the $\ads/\cft$ correspondence, the boundary CFT must contain a $\u(1)_{k}$ Kac-Moody current. Moreover, the value of the current $J_w$ is identified as the expectation value of the corresponding current operator $\hat{J}(w)$ in the dual CFT, and the component $A^{(0)}_{\bar{w}}$ plays the role of the source conjugate to the current operator. The CFT action is therefore deformed by an additional source term \cite{Kraus:2006wn}
	\begin{eqnarray}
		I_{CFT}\rightarrow I_{CFT}+\frac{i}{2\pi}\int dx^2 \sqrt{g^{(0)}} \hat{J}^{\bar{w}}A^{(0)}_{\bar{w}}.\label{souce to CFT}
	\end{eqnarray}
	Although the gauge fields decouple from gravity in the bulk, the additional boundary term couples to the metric and induces a shift for the stress tensor, which can be obtained by the variation of the action with respect to the boundary metric $g^{(0)}_{ij}$,
	\begin{eqnarray}
		\delta I_{A}=\frac{1}{2}\int_{\mathcal{\partial M}} dx^2 \sqrt{g^{(0)}} T^{ij}[A] \delta g^{(0)}_{ij}\ .\label{var action A}
	\end{eqnarray}
	Evaluating \eqref{var action A}, one finds
	\begin{eqnarray}
		T_{ww}[A]=\frac{k}{8\pi}A^{(0)}_{w}A^{(0)}_{w}\ ,\quad T_{\bar{w}\bar{w}}[A]=\frac{k}{8\pi}A^{(0)}_{\bar{w}}A^{(0)}_{\bar{w}}\ , \quad T_{w\bar{w}}[A]=T_{\bar{w}w}[A]=0\ .\label{A tensor}
	\end{eqnarray}
	Since $A^{(0)}$ is flat, $d A^{(0)}=0$, if we impose the gauge fixing such that $A^{(0)}_{w}$ is holomorphic and $A^{(0)}_{\bar{w}}$ is anti-holomorphic, the stress tensor \eqref{A tensor} is indeed conserved, $\nabla_{i}T^{ij}[A]=0$. However, it's well known that for a CFT with $\u(1)_{k}$ Kac-Moody current, the stress tensor arising from the current can be obtained by the Sugawara construction \cite{Sugawara:1967rw}
	\begin{eqnarray}
		\hat{T}_{J}(w)=\frac{1}{k}(\hat{J}\hat{J})(w) \, .
	\end{eqnarray}
	This suggests to modify the bulk stress tensor as\footnote[3]{In our conventions, the stress tensor on the gravity side is related to the CFT stress tensor by $T_{ww}=-\frac{1}{2\pi}T(w)$. This relation holds for both sectors, gravitational and gauge.}
	\begin{eqnarray}
		T_{ww}[J]&=&\frac{k}{8\pi}A^{(0)}_{w}A^{(0)}_{w}=-\frac{1}{2\pi}\frac{J(w)^2}{k}\ ,\nonumber\\ T_{\bar{w}\bar{w}}[J]&=& T_{w\bar{w}}[J]=T_{\bar{w}w}[J]=0\ .\label{TJ}
	\end{eqnarray}
	which is still conserved. Similarly, the anti-holomorphic part of the stess tensor is given by the right-moving field $\bar{A}$, i.e. $T_{\bar{w}\bar{w}}[\bar{J}]=-\frac{1}{2\pi}\frac{\bar{J}(\bar{w})^2}{k}$. Therefore, the non-vanishing components of total stress tensor read
	\begin{eqnarray}
		T_{ww}=T_{ww}[g]+T_{ww}[J]\ ,\quad T_{\bar{w}\bar{w}}=T_{\bar{w}\bar{w}}[g]+T_{\bar{w}\bar{w}}[\Bar{J}]\ .
	\end{eqnarray}
	Modes of the total stress tensor are defined as
	\begin{eqnarray}
		L_n-\frac{c}{24}\delta_{n,0}&=&-\oint dw\ e^{-inw}T_{ww}\ ,\nonumber\\
		\Bar{L}_n-\frac{c}{24}\delta_{n,0}&=&-\oint d\bar{w}\ e^{-in\bar{w}}T_{\bar{w}\bar{w}}\ .
	\end{eqnarray}
	As shown in \Appref{appendix A}, the full asymptotic symmetries imply the following quantum algebrae of the dual CFT,
	\begin{align}
		[J_{n}, J_m]&=\frac{1}{2}n k \delta_{m+n}\ ,\nonumber\\
		[L_{n}, J_m]&= -m J_{n+m}\ ,\nonumber\\
		[L_{n}, L_m]&=(n-m)L_{n+m}+\frac{c}{12}(n^3-n)\delta_{n+m,0}\ .\label{kac-vir}
	\end{align}
	which is the $\u(1)_k$ Kac-Moody-Virasoro algebra.
	
	\subsection{U(1) Wilson line defect in $\ads_3$}\label{section 3.2}
	In this section, we turn to investigate the gravity dual of the CFT charged moments $\mathcal{Z}_{1}(\mu)$, and we begin by introducing the notion of $U(1)$ subregion charge in the dual CFT. The Kac-Moody current $J_w$ in the dual CFT provides the left-moving charge,
	\begin{eqnarray}
		\hat{q}=\frac{1}{2\pi i}\oint dw \hat{J}_{w}\ .\label{q1}
	\end{eqnarray}
	On the gravity side, this charge generates the global $U(1)$ transformation for the gauge field (details are found in \Appref{appendix A}). Then the left-moving subregion charge on a boundary interval $\cA$ is obtained by replacing the contour integral in \eqref{q1} by the integration over $\cA$,
	\begin{eqnarray}
		\hat{q}_{\cA}=\frac{1}{2\pi i}\int_{\cA} dw \hat{J}_{w}\ .
	\end{eqnarray}
	Analogous relations hold for the right-moving subregion charge $\hat{\bar{q}}$. The total subregion charge is then given by
	\begin{eqnarray}
		\hat{Q}_{\cA}=\hat{q}_{\cA}+\hat{\bar{q}}_{\cA}\ .\label{sub-charge q}
	\end{eqnarray}
	
    \begin{figure}
	\begin{center}
		\begin{subfigure}[b]{0.3\textwidth}
			\includegraphics[width=\textwidth]{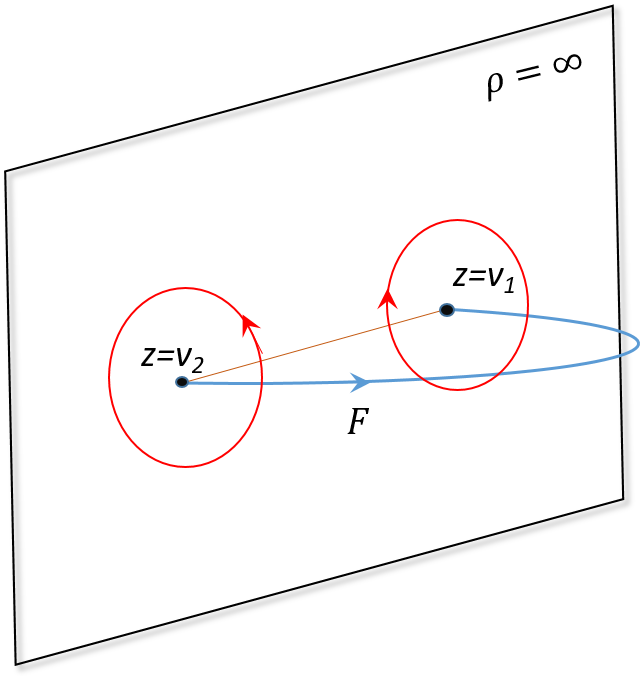}
			\caption{Defect in Poinca\'e $\ads_3$}
			\label{fig: Wilson line}
		\end{subfigure}
		\qquad\qquad
		\begin{subfigure}[b]{0.4\textwidth}
			\includegraphics[width=\textwidth]{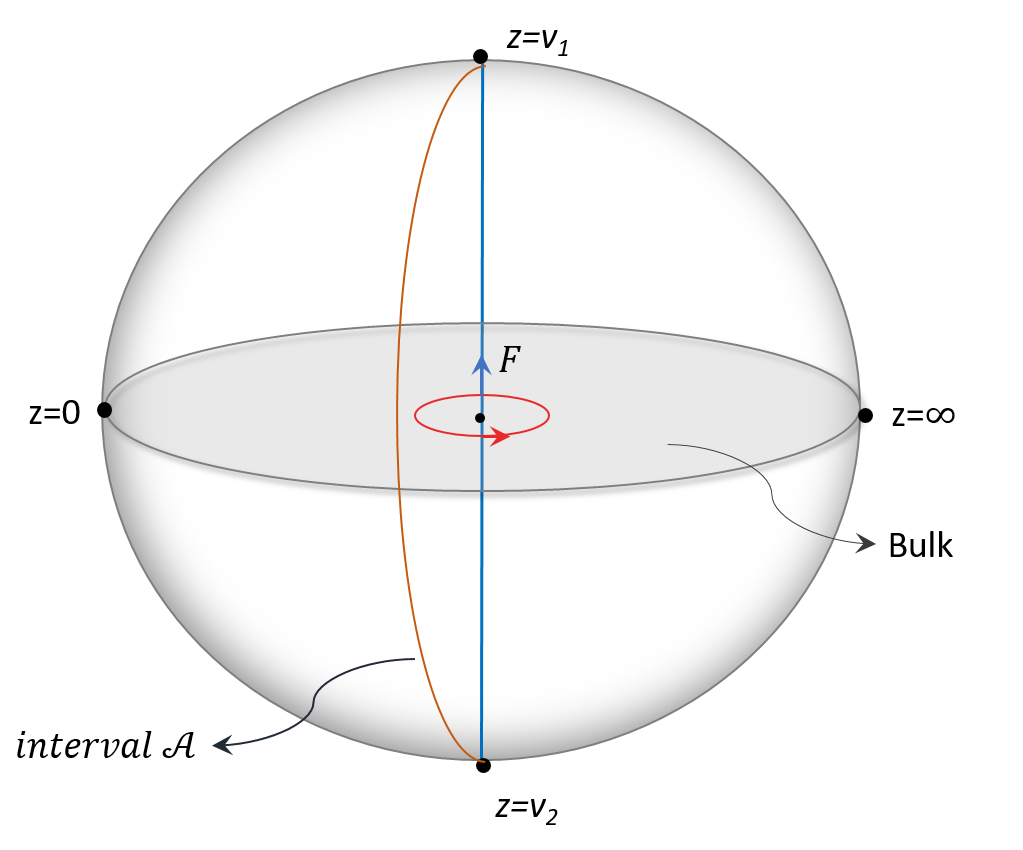}
			\caption{Riemann sphere}
			\label{fig: Z1 Riemann sphere}
		\end{subfigure}   
	\end{center}
	\caption{(a). In Poincar\'e $\ads_3$, the boundary defects extend continuously into the bulk as a Wilson line defect, inducing a non-trivial holonomy along the geodesic. (b). The endpoints of the boundary interval $\cA$ are transformed to the south pole and north pole of the Riemann sphere. The interior of the Riemann sphere represents the bulk, which contains a $U(1)$ symmetry when the background gauge fields vanish.}\label{fig: Z1 charged moments}
    \end{figure}

	The appearance of the non-local operator $e^{i\mu \hat{Q}_{\mathcal{A}}}$ in the charged moments indicates that on the gravity side, there should be an additional source coupled to the current which generates a shift for the original deformation of the CFT action. The question is what kind of source generates shifts of type $e^{i\mu \hat{Q}_{\cA}}$. Our inspiration is drawn from an observation in the dual CFT. As we show in \secref{section 5}, the operator $e^{i\mu Q_{\mathcal{A}}}$ acts as a defect in the dual CFT, inducing monodromies at the endpoints of the interval $\mathcal{A}$. Therefore, from the bulk point of view, it's natural that the defect located at the endpoints of $\mathcal{A}$ extends continuously into the bulk. This is synonymous with a Wilson line defect as illustrated in \figref{fig: Z1 charged moments}. Such a defect line induces non-trivial holonomy inside the bulk. As a consequence, it generates a phase for the gravitational path integral when the path encircles the defect. This parallels the CFT interpretation for the charged moments \cite{GoldsteinSela}, and is also in accordance with the cosmic brane interpretation of the holographic entanglement and R\'enyi entropy \cite{Lewkowycz:2013nqa, Ammon:2013hba, Dong:2016fnf}. Hence, to implement the this idea, we insert on the gravity side a defect action of form,
	\begin{eqnarray}\label{gravityDefect}
		I_{defect}=-\frac{ik\mu}{2\pi}\int_{\mathcal{C}}A-\bar{A}\ .\label{defect}
	\end{eqnarray}
	The curve $\mathcal{C}$ is anchored at the endpoints of the boundary interval $\mathcal{A}$. For the non-replica manifold, i.e. $n=1$, the trajectory of the curve is required to follow the path of the Ryu-Takayanagi geodesic. 
	
	\subsubsection{Euclidean Poincar\'e AdS$_{3}$}
	To test our holographic construction, we start with the Euclidean Poincar\'e $\ads_3$ background and reproduce the charged moments $\cZ_{1}(\mu)$ by solving the defect solutions. We focus on the left-movers $A$. The result for right-movers $\bar{A}$ is completely analogous. The line element of Poincar\'e $\ads_3$ reads
	\begin{eqnarray}
		ds^2&=&d\rho^2+e^{2\rho}dz d\bar{z},\quad i,j=1,2\ ,\label{Poincare}
	\end{eqnarray}
	with boundary complex coordinates defined as $z=x+it_E$ and $\bar{z}=x-it_E$. The ansatz for the background gauge field $A$ takes the asymptotic form
	\begin{align}
		A_z=a(z)+\cdots\ ,\quad A_{\bar{z}}=\bar{a}(\bar{z})+\cdots\ ,\quad \rho\to\infty\ .
	\end{align}
	Without loss of generality, we assume that the endpoints of the boundary interval $\mathcal{A}$ are located at $z=v_1$ and $z=v_2$, with $|v_1|\ge |v_2|$. After introducing the Wilson line defect \eqref{defect} in the bulk, the total action for the gauge field $A$ reads
	\begin{eqnarray}
		I=I_A+I_{d}=\frac{i k}{8\pi}\int_{\mathcal{M}} A\wedge d A-\frac{k}{16\pi}\int_{\mathcal{\partial M}} dx^2 \sqrt{g}A^{i}A_{i}-\frac{i k\mu}{2\pi}\int_{\mathcal{C}} ds A_{s}\ .
	\end{eqnarray}
	The curve follows the Ryu-Takayanagi geodesic, which is orthogonal to the boundary when approaching spatial infinity,  
	\begin{eqnarray}
		dx^{\mu}(s)\propto d\rho\ ,\quad \rho\to\infty\ .
	\end{eqnarray}
	The variation of the action with respect to $A$ gives
	\begin{eqnarray}
		\delta I = \frac{i k}{8\pi}\int dx^3 \delta A_{\mu}F_{\nu\sigma} \epsilon^{\mu\nu\sigma}-\frac{ik\mu}{2\pi}\int ds\int dx^3 \delta A_{\mu}\frac{d x^{\mu}}{ds}\delta^{(3)}\left(x-x(s)\right)+ \delta I_{bdy}\ .
	\end{eqnarray}
	with $\epsilon_{\mu\nu\sigma}$ being Levi-Civita tensor defined as $\epsilon_{\rho z \bar{z}}=\frac{i}{2}$.
	The equation of motion for $A$ becomes
	\begin{eqnarray}
		F_{\rho\lambda}=2\mu\int ds \frac{dx^{\mu}}{ds}\epsilon_{\mu\rho\lambda}\delta^{(3)}(x-x(s))\ .\label{eom2}
	\end{eqnarray}
	This equation of motion can be easily solved for a given curve. However, explicitely solving \eqref{eom2} is not necessary due to the following two properties which determine the asymptotic behavior of the gauge field. First, contracting \eqref{eom2} with the tangent vector to the curve, i.e. $dx^{\rho}/ds$, yields
	\begin{eqnarray}
		F_{\rho\lambda}\frac{dx^{\rho}}{ds}=2\mu\int ds \frac{dx^{\mu}}{ds}\frac{dx^{\rho}}{ds}\epsilon_{\mu\rho\lambda}\delta^{(3)}(x-x(s))=0\ .\label{condition1}
	\end{eqnarray}
	This implies that the non-vanishing components of field strength $F$ is tangent to the curve $\mathcal{C}$. Furthermore, integrating \eqref{eom2} on a two dimensional surface $\Sigma$ which intersects with $\mathcal{C}$ at an arbitrary point $x(s_0)$, one obtains
	\begin{eqnarray}
		\int_{\Sigma}F=\oint_{\partial\Sigma}A=2\mu\ .\label{condition2}
	\end{eqnarray}
	Since the curve is normal to the boundary when it approaches spatial infinity, from \eqref{condition1} and \eqref{condition2}, one finds
	\begin{eqnarray}
		\lim_{\rho\to\infty}F&=& F_{z\bar{z}}dz\wedge d\bar{z}\nonumber\\
		&=&2\mu\left(\delta^{(2)}(z-v_1,\bar{z}-\bar{v}_1)-\delta^{(2)}(z-v_2,\bar{z}-\bar{v}_2)\right)dz\wedge d\bar{z}\ .\label{field strength}
	\end{eqnarray}
	From \eqref{field strength}, the leading order of the gauge field solution is determined to be
	\begin{eqnarray}
		A^{(0)}_z= a(z)+B_z\ , \quad A^{(0)}_{\bar{z}}= \bar{a}(z)+B_{\bar{z}}\ ,\label{soultion A}
	\end{eqnarray}
	with the back-reaction contributions
	\begin{eqnarray}
		B_z=-\frac{i\mu}{2\pi}\left(\frac{1}{z-v_1}-\frac{1}{z-v_2}\right)\ , \quad B_{\bar{z}}=\frac{i\mu}{2\pi}\left(\frac{1}{\bar{z}-\bar{v}_1}-\frac{1}{\bar{z}-\bar{v}_2}\right)\ .\label{solution B}
	\end{eqnarray}
	There are residual gauge degrees of freedom for the back-reactions. However, as we verify in \secref{section 5}, if we consider the vacuum state with $a=\bar{a}=0$, the solution \eqref{solution B} matches the expectation value of the current operator.
	
	As mentioned in the previous section, the appearance of the source deforms the CFT action as in \eqref{souce to CFT}. In presence of the bulk Wilson line defect, \eqref{soultion A} and \eqref{solution B} imply a shift for the deformation of the CFT action
	\begin{eqnarray}\label{CFTshift}
		\Delta I_s=\frac{i}{2\pi}\int dx^2 \sqrt{g^{(0)}}\hat{J}^{\bar{z}}B_{\bar{z}}\ .\label{shift deformation}
	\end{eqnarray}
	Given the precise form of the source $B_{\bar{z}}$ \eqref{solution B}, the shift \eqref{CFTshift} can be further evaluated by expanding the source in Laurent series around $z=0$. We present this analysis only for the first term of $B_{\bar{z}}$ in \eqref{solution B},
	\begin{equation}
		B_{\bar{z}}^{(1)}=\frac{i\mu}{2\pi}\frac{1}{\bar{z}-\bar{v}_1}=\left\{
		\begin{aligned}
			-\frac{i\mu}{2\pi}\sum_{m=0}^{\infty}\bar{v}_{1}^{-m-1}\bar{z}^{m},\quad |z|<|v_1|\\
			\frac{i\mu}{2\pi}\sum_{m=0}^{\infty}\bar{v}_{1}^{m}\bar{z}^{-m-1},\quad |z|>|v_{1}|\,.\\
		\end{aligned}
		\right.\label{expansion source 1}
	\end{equation}
	The second pole in \eqref{solution B} is treated analogously. The mode expansion of the current operator $\hat{J}_z$ in complex $z$-plane is given by
	\begin{eqnarray}
		\hat{J}_{z}=\sum_{n=-\infty}^{\infty}z^{-n-1}J_{n}\label{expansion J z}\, .
	\end{eqnarray}
	In polar coordinates $z=\rho e^{i\theta}$ and $\bar{z}=\rho e^{-i\theta}$, the insertion of the expansions \eqref{expansion source 1} and \eqref{expansion J z} into \eqref{shift deformation} yields
	\begin{align}
		\Delta I^{(1)}_{s}&=\frac{i}{2\pi}\int dx^2 \sqrt{g^{(0)}} \hat{J}^{\bar{z}}B_{\bar{z}}^{(1)} \nonumber\\
		&=\frac{\mu}{2\pi^2}\int_{0}^{|v_1|}d\rho \ d\theta \  \sum_{n=-\infty}^{\infty}\sum_{m=0}^{\infty}J_{n} \rho^{m-n}e^{-i(n+1+m)\theta}\ \bar{v}_{1}^{-m-1}\nonumber\\
		&\ \ \ \ -\frac{\mu}{2\pi^2}\int_{|v_1|}^{\frac{1}{\epsilon}}d\rho \ d\theta\  \sum_{n=-\infty}^{\infty}\sum_{k=0}^{\infty}J_{n}\rho^{-n-k-1}e^{i(k-n)\theta}\ \bar{v}_{1}^{k}\nonumber\\
		%
		%
		&=-\frac{\mu}{2\pi}\sum_{n\neq 0}\frac{1}{n}J_{n}v_{1}^{-n}+\frac{\mu}{\pi}J_{0}\ln{|v_1\epsilon|}\label{eq}
	\end{align}
	where $\epsilon$ denotes the UV cut off of the integration, and we take the limit $\epsilon\to 0$ in the last step of \eqref{eq}. Following from \eqref{eq}, it is straightforward to write down the result for the full shift $\Delta I_{s}$,
	\begin{align}
		\Delta I_{s}&=-\frac{\mu}{2\pi}\sum_{n\neq 0}\frac{1}{n}J_{n}\left(v_{1}^{-n}-v_{2}^{-n}\right)+\frac{\mu}{\pi}J_{0}\ln{|\frac{v_1}{v_2}|}\ ,
	\end{align}
	which can be expressed in terms of the left-moving subregion charge operator \eqref{sub-charge q}, given by
	\begin{align}
		\Delta I_{s}=\frac{\mu}{2\pi}J_{0}\ln{\left(\frac{\bar{v}_1}{\bar{v}_{2}}\right)}+\frac{\mu}{2\pi}\int_{v_{2}}^{v_1}dz  \hat{J}_{z}= \frac{\mu}{2\pi}J_{0}\ln{\left(\frac{\bar{v}_1}{\bar{v}_{2}}\right)}-i\mu\hat{q}_{\mathcal{A}}\ .\label{eq2}
	\end{align}
	The analogous result for the right-moving sector is
	\begin{eqnarray}
		\Delta \Bar{I}_s=\frac{\mu}{2\pi}\bar{J}_{0}\ln{\left(\frac{v_1}{v_2}\right)}-i\mu\hat{\Bar{q}}_{\mathcal{A}}\ . \label{source to cft 2}
	\end{eqnarray}
	Combining \eqref{eq2} and \eqref{source to cft 2} yields the partition function in the dual CFT
	\begin{eqnarray}
		\langle e^{-\Delta I_{s}-\Delta\bar{I}_{s}}\rangle_{\text{gravity}}=\mathcal{Z}_{CFT}=\langle\  e^{i\mu \hat{Q}_{\mathcal{A}}-\Delta_0}\ \rangle_{CFT}\ ,\label{pt}
	\end{eqnarray}
	with
	\begin{eqnarray}\label{Delta0}
		\Delta_0=\frac{\mu}{2\pi}J_{0}\ln{\left(\frac{\bar{v}_1}{\bar{v}_{2}}\right)}+\frac{\mu}{2\pi}\bar{J}_{0}\ln{\left(\frac{v_1}{v_2}\right)}\,.
	\end{eqnarray}
	The partition function \eqref{pt} can be rewritten as
	\begin{eqnarray}
		\mathcal{Z}_{CFT}=\Tr{\left[\rho_{\mathcal{A}}\ e^{i\mu Q_{\mathcal{A}}-\Delta_0}\right]}\ ,\label{pt2}
	\end{eqnarray}
	which differs from the desired charged moments $\mathcal{Z}_{1}(\mu)=\Tr{\left[\rho_{\mathcal{A}}\ e^{i\mu \hat{Q}_{\mathcal{A}}}\right]}$ by the additional insertion of $e^{-\Delta_0}$. Since a $U(1)$ Wilson is the natural choice for the gravity dual of the charged moments, we expect that the exponential term $e^{-\Delta_0}$ can be removed by a proper holographic renormalization procedure. Due to our generating function method, c.f \secref{section 2.2}, we do not require the charged moments and hence not the properly renormalized action. Instead all that is needed is the correct expectation value of the subregion charge operator. In the case at hand this is given by
	\begin{eqnarray}
		\langle  \hat{Q}_{\mathcal{A}}\rangle_{1,\mu}=\frac{\langle  \hat{Q}_{\mathcal{A}} e^{i\mu \hat{Q}_{\mathcal{A}}}\rangle_{CFT}}{\langle e^{i\mu \hat{Q}_{\mathcal{A}}}\rangle_{CFT}}\,.
	\end{eqnarray}
	From the symmetry algebra \eqref{kac-vir}, we have $[\hat{J}_z, J_0]=0$, which implies the operator $e^{-\Delta_0}$ in \eqref{pt} doesn't carry charge. In other words, we have the following relation 
	\begin{eqnarray}
		\frac{\langle \hat{J}_z e^{i\mu \hat{Q}_{\mathcal{A}}}\rangle_{CFT}}{\langle e^{i\mu \hat{Q}_{\mathcal{A}}}\rangle_{CFT}}=\langle \hat{J}_z\rangle_{1,\mu}=\langle \hat{J}_z\rangle_{1,\mu,\Delta_0}=\frac{\langle \hat{J}_z e^{i\mu \hat{Q}_{\mathcal{A}}-\Delta_0}\rangle_{CFT}}{\langle e^{i\mu \hat{Q}_{\mathcal{A}}-\Delta_0}\rangle_{CFT}}\ .\label{J-eq}
	\end{eqnarray}
	The right-moving current $\hat{\bar{J}}_{\bar{z}}$ is treated in complete analogy. Therefore, the expectation value of the total subregion charge operator in these two different CFT states are the same, i.e.
	\begin{eqnarray}
		\langle  \hat{Q}_{\mathcal{A}}\rangle_{1,\mu}=\langle  \hat{Q}_{\mathcal{A}}\rangle_{1,\mu,\Delta_0}\ .\label{Q-eq}
	\end{eqnarray}
	Using the boundary values of the gauge field $A$ in \eqref{soultion A} and \eqref{solution B}, by the AdS/CFT correspondence, the left-moving subregion charge evaluates to
	\begin{align}
		\langle  \hat{q}_{\mathcal{A}}\rangle_{1,\mu}=\langle  \hat{q}_{\mathcal{A}}\rangle_{1,\mu,\Delta_0}
		&=\int_{v_2}^{v_1}\frac{dz}{2\pi i} \frac{ik}{2}a(z)+\int_{v_2+\epsilon}^{v_1-\epsilon}\frac{dz i}{2\pi} \frac{ik}{2}\cdot \left(-\frac{i\mu}{2\pi}\right)\left(\frac{1}{z-v_1}-\frac{1}{z-v_2}\right)\nonumber\\
		&=q_0+\frac{i k\mu}{4\pi^2}\ln{\left(\frac{v_1-v_2}{\epsilon}\right)}\ ,\label{charge value1}
	\end{align}
	where $q_0=\frac{k}{4\pi}\int_{v_2}^{v_1}dz\ a(z)$ denotes the background charge. In the second step of \eqref{charge value1}, since the integral is divergent, two cut-offs near the endpoints are introduced. The result for the right-moving charge takes an analogous form,
	\begin{eqnarray}
		\langle  \hat{\bar{q}}_{\mathcal{A}}\rangle_{1,\mu}=\bar{q}_0+\frac{i k\mu}{4\pi^2}\ln{\left(\frac{\bar{v}_1-\bar{v}_2}{\bar{\epsilon}}\right)}\ .
	\end{eqnarray}
	
	In the vacuum the background charge $q_0$ vanishes, and from \eqref{Q-f}, the normalized generating function $f_{1}(\mu)$ can be obtained as
	\begin{align}
		f_{1}(\mu)&=e^{i\int_{0}^{\mu} d\mu' \ \langle  \hat{q}_{\mathcal{A}}+\hat{\bar{q}}_{\mathcal{A}} \rangle_{1,\mu'}}=\left|\frac{v_1-v_2}{\epsilon}\right|^{-k\left(\frac{\mu}{2\pi}\right)^2}\ .\label{f_1}
	\end{align}
	As we verify in \secref{section 5}, this result coincides with the two-point function of the appropriately chosen charged vertex operators on the complex plane, where in particular the conformal dimensions of the vertex operators are given by
	\begin{eqnarray}
		\Delta_v=\bar{\Delta}_{v}=\frac{k}{4}\left(\frac{\mu}{2\pi}\right)^2
	\end{eqnarray}
	This agreement comprises a non-trivial check of the $\ads_3/\cft_2$ correspondence.
	
	\subsubsection{General asymptotic $\ads_3$ backgrounds}
	In this subsection, we generalize our holographic construction of the charged moments $\cZ_{1}(\mu)$ to asymptotic $\ads_3$ backgrounds. The most general solutions to three dimensional Euclidean Einstein gravity satisfying the Brown-Henneaux boundary conditions are the Ba\~nados geometries \cite{Banados:1998gg}. Their metric takes the form
	\begin{eqnarray}
		ds^2=d\rho^2+e^{2\rho}dw d\bar{w}+4G_3(\cL_g dw^2+\bar{\cL}_{g}d\bar{w}^2)+16G_3^2\cL_{g}\bar{\cL}_{g}e^{-2\rho}dw d\bar{w}\ ,\label{Banados metric}
	\end{eqnarray}
	where $\cL_{g}=\cL_{g}(w)$ and $\bar{\cL}_{g}=\bar{\cL}_{g}(\bar{w})$ are identified with the expectation values of the CFT stress tensor. One important feature of pure $\ads_3$ gravity is that all solutions are locally $\ads_3$. This feature allows to transform the Ba\~nados geometries to Poincar\'e $\ads_3$ by appropriate local coordinate transformations. In particular, in the case that $\cL_g(w)$ and $\bar{\cL}_g(\bar{w})$ are complex conjugates of each other, bulk diffeomorphisms correspond to associated conformal transformations for the CFT states. The explicit form of the bulk diffeomorphisms was found in \cite{Krasnov:2001cu}, which in fact is determined by the boundary conformal mapping. The stress tensor of Poinca\'re $\ads_3$ \eqref{Poincare} vanishes, $L_g(z)=0$, which implies the CFT state is the vacuum state on the complex plane. By the transformation law of the stress-tensor, the boundary conformal transformation from the $w$-coordinates to the flat $z$-plane satisfies
	\begin{eqnarray}
		\cL_g(w)=-\frac{c}{12}\{z,w\}\ ,\label{Sch de}
	\end{eqnarray}
	where $\{z,w\}$ denotes the Schwarzian derivative\footnote{The minus sign in \eqref{Sch de} appears since the imaginary part of $w=\phi+i\tau$ is Euclidean time. Note that the method presented in this section relies on $z(w)$ being bijective. If $z(w)$ defines a multi-sheeted Riemann surface, summing over images has to be taken into account, c.f. the example of conical defects in~Sec.~\ref{section 4.2}.}. Therefore, given $\cL(w)$, one can find the boundary conformal transformation $z=z(w)$ by solving the differential equation \eqref{Sch de}.
	
	One can use the conformal transformation $z=z(w)$ to obtain the asymptotic solution to the gauge field $A$ with the defect insertion. The reason is that the boundary condition $A^{(0)}_{\rho}=0$ is preserved under the asymptotic diffeomorphism \eqref{asy diffeo}, which implies that $A^{(0)}_{i}$ transforms as a vector under bulk asymptotic diffeomorphisms. Therefore, using \eqref{charge value1}, we directly obtain the leading order of the back-reaction in the asymptotic $\ads_3$ background as
	\begin{align}
		B_{w}&=-\frac{i\mu}{2\pi}\left(\frac{dz}{dw}\right)\left(\frac{1}{z(w)-z(w_1)}-\frac{1}{z(w)-z(w_2)}\right)\ ,\nonumber\\
		B_{\bar{w}}&=\frac{i\mu}{2\pi}\left(\frac{d\bar{z}}{d\bar{w}}\right)\left(\frac{1}{\bar{z}(\bar{w})-\bar{z}(\bar{w_1})}-\frac{1}{\bar{z}(\bar{w})-\bar{z}(\bar{w_2})}\right)\ ,\label{back-reaction z2}
	\end{align}
	where $w=w_1$ and $w=w_2$ denote the endpoints of the interval $\mathcal{A}$ on the $w$-plane. By inserting \eqref{back-reaction z2} into \eqref{shift deformation} and evaluating the integral explicitly, one finds that the shift \eqref{shift deformation} arising from the defect still takes the form of \eqref{eq2} in asymptotic $\ads_3$ backgrounds. In fact, there is a more convenient way to see this result. Since the deformation action \eqref{shift deformation} is conformally invariant, the shift for the deformation should take the same form as the Poincar\'e $AdS_3$ case, which in $w$-coordinates is given by
	\begin{align}
		\Delta I_s&=-\frac{\mu}{2\pi}\int_{\mathcal{A}}dw \hat{J}_{w}+\frac{\mu}{2\pi}J_{0}\ln{\left(\frac{\bar{z}(\bar{w}_1)}{\bar{z}(\bar{w}_2)}\right)}=\frac{\mu}{2\pi}J_{0}\ln{\left(\frac{\bar{z}(\bar{w}_1)}{\bar{z}(\bar{w}_2)}\right)}-i\mu \hat{q}_{\mathcal{A}}\ .\label{source general}
	\end{align}
	Again, this implies that the expectation value of the subregion charge for the charged moments $\cZ_(\mu)$ is the same as the boundary current of the gauge field. To derive the left-moving subregion charge from the boundary value of the gauge field, we notice that $q_{\mathcal{A}}$ is invariant under conformal mappings. It is more convenient to introduce an additional global conformal mapping, i.e.
	\begin{eqnarray}
		z\to z'=\frac{z-z(w_2)}{z-z(w_1)}\ ,
	\end{eqnarray}
	which maps the $z=z(w_2)$ to $z'=0$ and $z=z(w_1)$ to $z'=\infty$. Then the corresponding back-reaction takes the simple form
	\begin{eqnarray}
		B_{z'}=\frac{i\mu}{2\pi z'}\ ,\quad B_{\bar{z}'}=-\frac{i\mu}{2\pi \bar{z}'}\ .
	\end{eqnarray}
	Working in $z'$-coordinates, we find that the subregion charge takes the following universal form,
	\begin{align}
		\langle \hat{q}_{\mathcal{A}}\rangle_{1,\mu}=\langle \hat{q}_{\mathcal{A}}\rangle_{1,\mu,\Delta_0}&=q_0+\frac{ik}{2}\int_{\epsilon(\delta)}^{\frac{1}{\epsilon(\delta)}}\frac{dz' }{2\pi i} B_{z'}=q_0+\frac{i k\mu}{4\pi^2}\ln{\left(\frac{1}{\epsilon(\delta)}\right)}\ ,\label{charge general}
	\end{align}
	where $q_0$ is the background charge, and $\epsilon(\delta)$ denotes the cut-off on the $z'$-plane, induced by the corresponding cut-off $\delta$ on the $w$-plane through the conformal transformation $z'=z'(z(w))$,
	\begin{eqnarray}
		\epsilon(\delta)=z'\left(z(w_2+\delta)\right)-z'\left(z(w_2)\right)=z'\left(z(w_2+\delta)\right)\ .
	\end{eqnarray}
	The right-moving subregion charge takes a similar form
	\begin{eqnarray}
		\langle \hat{\bar{q}}_{\mathcal{A}}\rangle_{1,\mu}=\langle \hat{\bar{q}}_{\mathcal{A}}\rangle_{1,\mu,\Delta_0}=\bar{q}_0+\frac{i k\mu}{4\pi^2}\ln{\left(\frac{1}{\bar{\epsilon}(\bar{\delta})}\right)}\ , \label{charge general 2}
	\end{eqnarray}
	where $\bar{q}_0$ denotes the right-moving background charge. The anti-holomorphic cut-offs $\bar{\epsilon}$ and $\bar{\delta}$ are the complex conjugates of $\epsilon$ and $\delta$. Combining \eqref{charge general} and \eqref{charge general 2} yields the total subregion charge for the charged moments $\cZ_{1}(\mu)$, given by
	\begin{align}
		\langle \hat{Q}_{\mathcal{A}}\rangle_{1,\mu}=\langle \hat{q}_{\mathcal{A}}\rangle_{1,\mu}+\langle i\hat{\bar{q}}_{\mathcal{A}}\rangle_{1,\mu}=Q_0+\frac{i k\mu}{2\pi^2}\ln{\left|\frac{1}{\epsilon(\delta)}\right|}\ ,\label{charge total 1}
	\end{align}
	where $Q_0=q_0+\bar{q_0}$ denotes the total background charge. In particular, as we show in \secref{section 4}, the logarithm in \eqref{charge total 1} is related to the length of the geodesic anchored at the endpoints of the interval in the background geometry.
	
	\subsection{Subregion charge in the replica manifold}\label{section 3.3}
	In this subsection, we apply our holographic construction of charged moments $\cZ_{n}(\mu)$ to the replica manifold. In the replica manifold, the $U(1)$ Wilson line defect \eqref{defect} considered before must follow the $\Z_n$ fixed points. From the Chern-Simons gauge field sourced by the Wilson line,  we calculate the subregion charge, which will be used in the derivation of the results of \secref{section 4}.
	
	\begin{figure}
		\begin{center}
		   \includegraphics[scale=0.4]{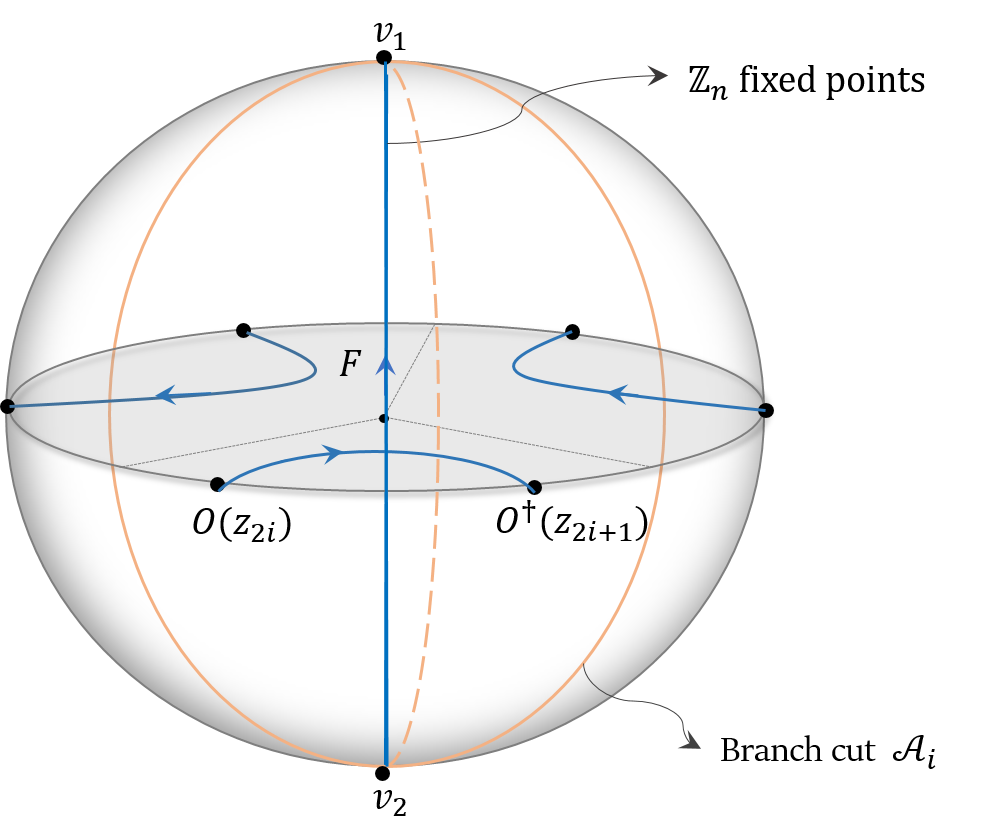}
		\end{center}
		\caption{For the complex plane with a single interval as the entanglement region, the corresponding $n$-sheeted branched covering is still a Riemann sphere. The defect in the bulk replica manifold runs along the locus of the $\Z_n$ fixed points. The operators $O$ denote the current primary fields for the original CFT states, which generate the background gauge fields. They are taken to be $n$ copies on the replica manifold.}\label{fig: Zn charged moments}
	\end{figure}
	
	For pure $\ads_3$ gravity, the exact replica manifold $\cM_n$ was found in \cite{Hartman:2013mia}, using the Schottky uniformization method. A characteristic feature of these geometries is that they contain a $\Z_{n}$ symmetry in the bulk. The corresponding line of $\Z_{n}$ fixed points is located along the geodesic anchored at the boundary branch points. This $\Z_{n}$ symmetry persists when matter fields exist in the bulk. On the boundary $n$-sheeted Riemann surface $\cR_n$, there are $n$ branch cuts lying on the $n$ separate sheets. As illustrated in \figref{fig: Zn charged moments}, our $U(1)$ Wilson line defect must run along these $\Z_n$ fixed points in order to preserve the $\Z_n$ symmetry.\footnote{Technically speaking, the operator $e^{i\mu\hat{Q}_{\cA}}$ in the charged moments $\cZ_{n}(\mu)$ is defined on a single sheet. When using the replica trick, this operator should be evenly distributed on $n$ branch cuts of $\cR_n$. This construction ensures that the charged moments preserves the $\Z_{n}$ symmetry on the replica manifold.} 
	
	For simplicity, consider the Poinca\'e $\ads_3$ with $\cA=[v_2, v_1]$ as the entangling interval on the boundary $\cR_1$. As shown in \cite{Hung:2011nu}, the corresponding replica solution assumes the form of the Ba\~nados geometries \eqref{Banados metric}. The complex coordinates on the spatial boundary $\cR_1$ of Poincar\'e $\ads_3$, denoted with $z$ and $\bar{z}$, are analytically continued to the $n$-sheeted branch covering $\cR_n$. The line element on $\cR_n$ can be written as
	\begin{eqnarray}
		d\Tilde{s}^2=g^{(0)}_{i j}dx^i dx^j =dz d\Bar{z}\ .
	\end{eqnarray}
	One should note that in fact the metric is singular at the endpoints of $\cA$ since the range of angle around the endpoints is $\Delta \theta=2\pi n$. The curvature singularities at the endpoints can be extracted from the Weyl factor of the following conformal mapping from the universal cover of the $z$-plane to a flat $z'$-complex plane \cite{Calabrese:2004eu}
	\begin{eqnarray}
		z'=\left(\frac{z-v_2}{z-v_1}\right)^{\frac{1}{n}}\ ,\label{mapping 1}
	\end{eqnarray}
	where the endpoints $z=v_2$ and $z=v_1$ are mapped to $z'=0$ and $z'=\infty$. By this conformal mapping, the states in the dual CFT are transformed to the vacuum state on the complex $z'$-plane. The full bulk diffeomorphisms to the Poincar\'e $\ads_3$ were given in \cite{Hung:2011nu}. 
	
	To compute the subregion charge expectation value on the gravity side, we assume that the components of the background gauge field $A$ on the original manifold $\mathcal{M}_1$ take the following boundary value,
	\begin{eqnarray}
		A^{(0)}_{z}=a(z)\ ,\quad \quad A^{(0)}_{\bar{z}}=\bar{a}(\bar{z})\ .
	\end{eqnarray}
	As mentioned in \secref{section 3.2}, the boundary value of the gauge field transforms as a tensor under the conformal transformation. In presence of the $U(1)$ Wilson line defect, using the conformal mapping \eqref{mapping 1} and  the equation \eqref{back-reaction z2}, a straightforward calculation yields the boundary value of the gauge field
	\begin{eqnarray}
		A^{(0)}_{z}=a(z)+B_z\ ,\quad \quad A^{(0)}_{\bar{z}}=\bar{a}(\bar{z})+B_{\bar{z}}\ ,
	\end{eqnarray}
	with the back-reaction
	\begin{eqnarray}
		B_z=-\frac{i\mu}{2\pi n}\left(\frac{1}{z-v_1}-\frac{1}{z-v_2}\right)\ , \quad B_{\bar{z}}=\frac{i\mu}{2\pi n}\left(\frac{1}{\bar{z}-\bar{v}_1}-\frac{1}{\bar{z}-\bar{v}_2}\right)\ .\label{solution B 3u}
	\end{eqnarray}
	One can check that, inserting the solution \eqref{solution B 3u} into \eqref{shift deformation} indeed yields a shift of the deformation of the form of \eqref{source general}. The boundary left-moving subregion charge is obtained as
	\begin{align}
		\langle \hat{q}_{\mathcal{A}}\rangle_{n,\mu}&=\frac{ik}{2}\int_{v_2}^{v_1}\frac{dz}{2\pi i} A^{(0)}_z=q_0+\frac{i k\mu}{4\pi^2 n}\ln{(\frac{v_1-v_2}{\delta})}\ ,\label{n-charge 1}
	\end{align}
	where $q_0$ denotes the background charge, and $\delta$ is the cut-off near the end points $v_1$ and $v_2$ on $\mathcal{R}_n$. One can confirm this result via the general formula \eqref{charge general} and the mapping \eqref{mapping 1}. In this case, the cut-off $\epsilon(\delta)$ in \eqref{charge general} is given by
	\begin{eqnarray}
		\epsilon(\delta)=\left(\frac{\delta}{v_1-v_2}\right)^{\frac{1}{n}}\, .
	\end{eqnarray}
	The analogous results can be obtained for the right-moving subregion charge. Hence, the total charge \eqref{sub-charge q} is
	\begin{eqnarray}\label{total charge n}
		\langle \hat{Q}_{\mathcal{A}}\rangle_{n,\mu}=Q_0+\frac{i k\mu}{2\pi^2 n}\ln{\left|\frac{v_1-v_2}{\delta}\right|}\ ,
	\end{eqnarray}
	The subregion charge differs from the results for $n=1$ by the $\frac{1}{n}$ factor. This implies that the charge is evenly distributed on the $n$ sheets of $\cR_n$, which preserves the $\Z_n$ symmetry of the bulk solutions. For the replica manifolds of the general $AdS_3$ solution, one can always transform to the Poincar\'e case and compute the charge by the same procedure.
	
	\section{Holographic calculation}\label{section 4}
	In this section, we consider two examples of background solutions and calculate the subregion charges for the replica manifolds with defect insertions. We show that the probability distributions $P_{\mathcal{A},n}(q)$ are the Gaussian distributions of the charge fluctuations in the general cases. Furthermore, the resulting symmetry-resolved entanglement entropy is independent of the subregion charge and completely determined by the length of the geodesic in the background solutions and the Chern-Simons level.
	\subsection{Example 1: Poincar\'e $\ads_3$}\label{section 4.1}
	Using Poincar\'e $AdS_3$,
	\begin{eqnarray}
		ds^2=d\rho^2+e^{2\rho}dz d\bar{z}\ ,
	\end{eqnarray}
	as background geometry, the background gauge fields have the general asymptotic expansion
	\begin{eqnarray}
		A_{z}=a_1(z)+\cdots\ ,\quad \quad A_{\bar{z}}=\bar{a}_1(\bar{z})+\cdots\ ,\nonumber\\
		\bar{A}_{z}=a_2(z)+\cdots\ ,\quad \quad A_{\bar{z}}=\bar{a}_2(\bar{z})+\cdots\ .\label{Charged Poincare}
	\end{eqnarray}
	Given a boundary interval with the endpoints located at $z=v_1$ and $z=v_2$, \eqref{total charge n} readily provides the expectation value total subregion charge for the charged moments, denoted $\langle Q_{\mathcal{A}}\rangle_{n,\mu}$. Then the normalized generating function is obtained through
	\begin{align}
		f_{n}(\mu)&=\exp{\left(i\int_{0}^{\mu} d\mu'\ \langle \hat{Q}_{\mathcal{A}}\rangle_{n,\mu'}\right)}
		=
		e^{i\mu Q_0}\cdot\left|\frac{v_1-v_2}{\delta}\right|^{-\frac{k}{n}(\frac{\mu}{2\pi})^2}\,,
	\end{align}
	where $Q_0=q_0+\bar{q}_0$ is the total subregion charge stemming purely from the background solutions \eqref{Charged Poincare}.
	In Poincar\'e AdS, the regularized length of the geodesic which connects the endpoints of the boundary interval $\mathcal{A}$ is given by
	\begin{eqnarray}
		L=2\ln{\left|\frac{v_1-v_2}{\delta}\right|}\,,
	\end{eqnarray}
	where $\delta$ is related to the bulk UV-cutoff $\rho_0$, i.e. $\delta=e^{-\rho_0}$, and should be identified with the cutoff around the boundary endpoints. The subregion charge and the normalized generating function are then expressed through
	\begin{eqnarray}
		\langle \hat{Q}_{\mathcal{A}}\rangle_{n,\mu}=Q_0+\frac{i k\mu}{4\pi^2 n}L\ ,\quad f_{n}(\mu)=e^{i\mu Q_0-\frac{k}{2n}(\frac{\mu}{2\pi})^2 L}\ . \label{ex 1 Q f}
	\end{eqnarray}
	The Fourier transformation of $f_{n}(\mu)$ in \eqref{ex 1 Q f} yields the probability distribution $P_{\mathcal{A},n}(q)$
	\begin{eqnarray}\label{Pdistribution}
		P_{\mathcal{A},n}(q)=\int_{-\infty}^{\infty}\frac{d\mu}{2\pi} e^{-i\mu q}f_{n}(\mu)=\sqrt{\frac{2\pi n}{k L}}e^{-\frac{2n\pi^2\Delta q^2}{k L}}\,,
	\end{eqnarray}
	where the fluctuation of the subregion charge is denoted by $\Delta q=q-Q_0$.
	The resulting symmetry-resolved entanglement entropy \eqref{SEE simp} is then easily obtained as
	\begin{eqnarray}\label{chargedAdSSREE}
		S(q)=S+\lim_{n\to 1}\frac{1}{1-n}\log{\frac{P_{\mathcal{A},n}(q)}{P_{\mathcal{A}}(q)^n}}=\frac{c}{6}L-\frac{1}{2}\ln{\left(\frac{kL}{2\pi}\right)}+O(1)\ ,
	\end{eqnarray}
	where the entanglement entropy $S$ is given by the Ryu-Takayanagi formula in three dimensions, $S=\frac{c}{6}L$ in three dimensions \cite{RT}. Observe that the charge dependence has disappeared in the final result. This is called equipartition of the entanglement entropy \cite{Xavier:2018kqb} and indicates that all charge sector have identical entanglement.
	
	\subsection{Example 2: Global $\ads_3$ and conical defect}\label{section 4.2}
	Asymptotic $\ads_3$ geometries at zero temperature and compact spatial coordinate, $\phi\sim\phi+2\pi$ are described by the line element
	\begin{eqnarray}
		ds^2=\frac{1}{\frac{1}{N^2}+r^2}dr^2+(\frac{1}{N^2}+r^2)d\tau^2+r^2 d\phi^2\ ,\label{global AdS 1}
	\end{eqnarray}
	These are parametrized by a positive real number $N$. For $N=1$, the line element is global $AdS_3$ while for $N>1$, it corresponds to the conical defects with deficit angle  $2\pi(1-\frac{1}{N})$ at $r=0$.
	
	In new coordinates
	\begin{eqnarray}
		r=e^{\rho}-\frac{1}{4N}e^{-\rho}\ , \quad w=\phi+i\tau\ ,\quad \bar{w}=\phi-i\tau\,,
	\end{eqnarray}
	the line element \eqref{global AdS 1} assumes the form of the Ba\~nados geometries \eqref{Banados metric} with
	\begin{eqnarray}
		\cL_g=\bar{\cL}_g=-\frac{1}{16G_3N^2}=-\frac{c}{24N^2}\ .\label{Lg1}
	\end{eqnarray}
	By the local coordinate transformations
	\begin{eqnarray}
		z=\frac{r}{\sqrt{r^2+\frac{1}{N^2}}}e^{(\tau-i\phi)/N},\quad \xi=\ln{\sqrt{N^2 r^2+1}}-\frac{\tau}{N}\ ,\label{AdS-poincare}
	\end{eqnarray}
	the line element \eqref{global AdS 1} is transformed to Poincar\'e $AdS_3$,
	\begin{eqnarray}
		ds^2=d\xi^2+e^{2\xi}dz d\bar{z}\ .
	\end{eqnarray}
	On the boundary, the coordinate transformations \eqref{AdS-poincare} reduce to the conformal mapping,
	\begin{eqnarray}
		z=e^{(\tau-i\phi)/N}=e^{-iw/N}\ ,\label{mapping conical}
	\end{eqnarray}
	which indeed reproduces the stress tensor \eqref{Lg1} by the conformal  transformation law \eqref{Sch de}. Since the mapping \eqref{mapping conical} between the conical defect and the covering space-time is not bijective, inserting a Wilson line defect in the conical defect $\ads_3$ leads to $N$ copies of that Wilson line defect in the covering Poincar\'e $\ads_3$ space-time. Assuming a spatial boundary interval $\cA$ with endpoints $w_1=i\tau_0+\phi_1$ and $w_2=i\tau_0+\phi_2$, the locations of the $2N$ endpoints of the defects in the $z$-plane are given by
	\begin{eqnarray}
		z_{2j-1}=e^{-i(w_1+2\pi j)/N}\ ,\quad z_{2j}=e^{-i(w_2+2\pi j)/N}\ ,\quad j=1, 2,\cdots, N\ ,
	\end{eqnarray}
	which leads to the asymptotic form of back-reaction in the $z$-plane as
	\begin{align}
		B_{z}&=-\frac{i\mu}{2\pi}\sum_{j=1}^{N}\left(\frac{1}{z-z_{2j-1}}-\frac{1}{z-z_{2j}}\right)\ ,\nonumber \\
		B_{\bar{z}}&=\frac{i\mu}{2\pi}\sum_{j=1}^{N}\left(\frac{1}{\bar{z}-\bar{z}_{2j-1}}-\frac{1}{\bar{z}-\bar{z}_{2j}}\right)\ .
	\end{align}
	Hence, by the transformation \eqref{mapping conical}, one finds the solutions for the back-reaction in $w$-coordinates to be
	\begin{align}
		B_{w}&=\frac{\mu}{2\pi}\left(\frac{e^{i w_1}}{e^{i w}-e^{i w_1}}-\frac{e^{i w_2}}{e^{i w}-e^{i w_2}}\right)\ ,\nonumber\\
		B_{\bar{w}}&=\frac{\mu}{2\pi}\left(\frac{e^{-i \bar{w}_1}}{e^{-i \bar{w}}-e^{-i \bar{w}_1}}-\frac{e^{-i \bar{w}_2}}{e^{-i \bar{w}}-e^{-i \bar{w}_2}}\right)\ .\label{back-reaction conical}
	\end{align}
	Note that the solutions \eqref{back-reaction conical} are independent of $N$, owing to the fact that the $U(1)$ Chern-Simons fields are decoupled from gravity. In the $n$-replica manifold with branch cut $\mathcal{A}$ on the boundary, the back-reaction of the Wilson line defect reads
	\begin{align}
		B_{w}&=\frac{\mu}{2\pi n}\left(\frac{e^{i w_1}}{e^{i w}-e^{i w_1}}-\frac{e^{i w_2}}{e^{i w}-e^{i w_2}}\right)\ ,\nonumber\\
		B_{\bar{w}}&=\frac{\mu}{2\pi n}\left(\frac{e^{-i \bar{w}_1}}{e^{-i \bar{w}}-e^{-i \bar{w}_1}}-\frac{e^{-i \bar{w}_2}}{e^{-i \bar{w}}-e^{-i \bar{w}_2}}\right)\ .\label{back-reaction conical replica}
	\end{align}
	The additional $\frac{1}{n}$ factor appearing in \eqref{back-reaction conical replica} yields the correct holonomy when each endpoint is circled $n$ times.
	
Hello Analogous to the Poincar\'e patch case discussed in Sec.~\ref{section 4.1}, general on-shell solutions are allowed for the background gauge fields. The only requirement is that they satisfy the boundary condition $A_\rho=0$, i.e. the analog of \eqref{Charged Poincare}. Since in this case, the spatial direction is compact, general background solutions might be singular in the interior. However, the singular behavior can be valid if $U(1)$ Wilson defects exist in the background solution. By the AdS/CFT correspondence, these are generated by appropriate boundary current primary operators located at the endpoints of the defect. Hence, as before, we simply denote the total background subregion charge as $Q_0=q_0+\bar{q}_0$. The final result for the subregion charge of the $n$-replica solutions in presence of the defect is given by
	\begin{align}
		\langle \hat{Q}_{\mathcal{A}}\rangle_{n,\mu}=Q_0+\frac{i k\mu}{2\pi^2 n}\ln{\left|\frac{2}{\delta}\sin{\frac{\Delta\phi}{2}}\right|}\ ,\label{charge global n}
	\end{align}
	with $\Delta\phi=\phi_1-\phi_2$. $\delta$ denotes the cut-off around the endpoints in the $w$-coordinates.
	
	Note that the regularized length of the geodesic in global $AdS_3$ or in the conical defect geometry reads
	\begin{eqnarray}
		L_{N}=2\ln{\left|\frac{2N}{\delta}\sin{\frac{\Delta\phi}{2N}}\right|}\ .\label{geo length 2}
	\end{eqnarray}
	As a result, we find that the subregion charge and the normalized generating function take the same form as in the Poincar\'e case \eqref{ex 1 Q f}, with the only difference being the exact form of the geodesic length. Hence, the result for the symmetry-resolved entanglement entropy is given by
	\begin{eqnarray}
		S(q)=\frac{c}{6}L_N-\frac{1}{2}\ln{\left(\frac{kL_1}{2\pi}\right)}+O(1)\ .
	\end{eqnarray}
	Again, we find equipartition of the entanglement entropy. This concludes our gravity analysis.
	
	\section{CFT calculation}\label{section 5}
	In this section, we provide a CFT calculation of the symmetry-resolved entanglement entropy for the vacuum state defined on the complex plane. The result is consistent with our holographic calculation for the uncharged Poincar\'e $AdS_3$ background, cf \secref{section 4.1}.
	\subsection{Non-local operator from vertex operators}
	In this subsection, we show how to express the $U(1)$ Wilson line defect through a pair of bosonic vertex operators. This simplifies the anaylsis, since vertex operators have the desirable property of being $\u(1)_k$ primary (which includes Virasoro primarity). Furthermore, it explains the appearance of vertex operators in \cite{GoldsteinSela} from our defect vantage point.
	
	As demonstrated in the previous section, the dual CFT of $U(1)$-Chern-Simons Einstein gravity contains a $\u(1)_{k}$ Kac-Moody current $\hat{J}(z)$. The symmetry algebra is now given by
	\begin{eqnarray}
		\left[J_{n}, J_m\right]&=&\frac{1}{2}n k \delta_{m+n,0} \ ,\nonumber\\
		\left[L_n, J_m\right]&=& -m J_{n+m} \ ,\nonumber\\
		\left[L_n, L_m\right]&=&(n-m)L_{n+m}+\frac{c}{12}(n^3-n)\delta_{n+m,0}\ ,
	\end{eqnarray}
	and the corresponding OPEs read
	\begin{eqnarray}
		\hat{J}(z)\hat{J}(w)&\sim & \frac{k/2}{(z-w)^2}+\cdots,\nonumber\\
		\hat{T}(z)\hat{J}(w)&\sim & \frac{\hat{J}(w)}{(z-w)^2}+\frac{\partial \hat{J}(w)}{z-w}+\cdots,\nonumber\\
		\hat{T}(z)\hat{T}(w)&\sim & \frac{c/2}{(z-w)^4}+\frac{2\hat{T}(w)}{(z-w)^2}+\frac{\partial \hat{T}(w)}{z-w}+\cdots \ .\label{OPE1}
	\end{eqnarray}
	For the right-moving part, analogous relations hold. The first task for the CFT calculations is to investigate the properties of the non-local operator $e^{i\mu \hat{Q}_{\mathcal{A}}}$. Its holomorphic part given by
	\begin{eqnarray}
		I_{\mu}(v_1, v_2) := e^{i\mu\hat{q}_{\mathcal{A}}} =\exp{\left(\frac{\mu}{2\pi}\int_{v_2}^{v_1}dz \hat{J}(z)\right)}\ ,\label{hol non-local I}
	\end{eqnarray}
	where the positions of the endpoints are arranged such that $|v_1|\ge|v_2|$. This non-local operator should be radially ordered in correlation functions. As we show in Appendix \ref{appendix B}, one can rewrite the radially ordered non-local operator as
	\begin{eqnarray}
		\hat{R}\left[I_{\mu}(v_1, v_2)\right]=\left(\frac{v_1-v_2}{\delta}\right)^{-\frac{k}{2}(\frac{\mu}{2\pi})^2}  \normord{I_{\mu}(v_1, v_2)}\  ,\label{id 1}
	\end{eqnarray}
	where $\hat{R}$ denotes the radial ordering operator, and the cut-offs $\delta$ are introduced around the endpoints in order to regulate the non-local operator $I_{\mu}(v_1, v_2)$. On the other hand, if we formally write $\chi(z)=\int^{z}dw \hat{J}(w)$, a vertex operator is expressed through
	\begin{eqnarray}
		V_{\mu}(z)=\normord{\exp{\left(\frac{\mu}{2\pi}\chi(z)\right)}}\ .
	\end{eqnarray}
	Then the following identity holds in any correlation function \cite{GaberdielKlemmRunkel},
	\begin{eqnarray}
		V_{\mu}(v_1)V_{-\mu}(v_2)=(v_1-v_2)^{-\frac{k}{2}(\frac{\mu}{2\pi})^2}\normord{I_{\mu}(v_1, v_2)} \ .\label{id 2}
	\end{eqnarray}
	Comparing \eqref{id 1} with \eqref{id 2} leads to the following relation,
	\begin{eqnarray}
		I_{\mu}(v_1, v_2)=\delta^{\frac{k}{2}(\frac{\mu}{2\pi})^2} V_{\mu}(v_1)V_{-\mu}(v_2)\ ,
	\end{eqnarray}
	where the cut-off $\delta$ can be regarded as a natural normalization constant rendering the two point function of the vertex operators dimensionless. In the following we identify the operator $I_{\mu}(v_1, v_2)$ with the pair of vertex operators $V_{\mu}(v_1)V_{-\mu}(v_2)$, since the cut-off dependent prefactor cancels in all ratios of correlators of interest. Thus, in the following the focus lies on the analysis of vertex operators. 
	
	As shown
	\footnote{More precisely, in \appref{appendix B} we rederive the familiar $\u(1)$ OPEs \eqref{U1OPE} without assuming the Sugawara construction in order to minimize the employed assumptions. Of course the result confirms that the energy momentum tensor must be of Sugawara form.}
	in \appref{appendix B}, the OPEs for the vertex operator are given by
	\begin{align}\label{U1OPE}
		\hat{J}(z)V_{\mu}(0)&\sim\frac{k}{2}\left(\frac{\mu}{2\pi}\right)\frac{V_{\mu}(0)}{z}+\cdots\nonumber\\
		\hat{T}(z)V_{\mu}(0)&\sim\frac{k}{4}\left(\frac{\mu}{2\pi}\right)^2\frac{V_{\mu}(0)}{z^2}+\frac{\partial V_{\mu}(0)}{z}+\cdots\ .
	\end{align}
	Thus, combining with the anti-holomorphic part, the full vertex operator $V_{\mu}(z, \bar{z})=V_{\mu}(z)\bar{V}_{\mu}(\bar{z})$ are both Virasoro primary and $\u(1)$ primary, with conformal dimensions and charges given by
	\begin{eqnarray}\label{confWeightCharge}
		\Delta_{\mu}=\bar{\Delta}_{\mu}=\frac{k}{4}\left(\frac{\mu}{2\pi}\right)^2,
		\quad
		\charge_\mu=\bar{\charge}_{\mu}=\frac{k}{2}\left(\frac{\mu}{2\pi}\right)\ .
	\end{eqnarray}
	With this, the vacuum expectation value of the current in presence of the vertex operators is given by
	\begin{eqnarray}
		\frac{\langle \hat{J}(w)V_{\mu}(v_1, \bar{v}_1)V_{-\mu}(v_2, \bar{v}_2)\rangle}{\langle V_{\mu}(v_1, \bar{v}_1)V_{-\mu}(v_2, \bar{v}_2)\rangle}
		=
		\frac{k}{2}\left(\frac{\mu}{2\pi}\right)\left(\frac{1}{w-v_1}-\frac{1}{w-v_2}\right)\ ,
	\end{eqnarray}
	which indeed coincides with the solutions to the gauge field \eqref{soultion A} in presence of the Wilson line defect \eqref{defect} for the uncharged Poincar\'e $AdS_3$ background. On the other hand, the two-point function of vertex operators reads
	\begin{eqnarray}
		\langle V_{\mu}(v_1, \bar{v}_1)V_{-\mu}(v_2, \bar{v}_2)\rangle=\left|\frac{v_1-v_2}{\delta}\right|^{-k(\frac{\mu}{2\pi})^2}\ ,
	\end{eqnarray}
	which agrees with the normalized generating function $f_{1}(\mu)$ in \eqref{f_1} for the uncharged Poincar\'e $AdS_3$.
	
	\subsection{Symmetry-resolved entanglement through charged twist operators}
	In this section, following \cite{GoldsteinSela}, we use the charged twist operators for the computation of the subregion charge expectation value $\langle \hat{Q}_\cA\rangle_{n,\mu}$, the charged moments and finally, via our new generating function, the symmetry-resolved entanglement entropy.  The resulting symmetry-resolved entanglement entropy coincides with the holographic result of \secref{section 4.1}. Our analysis extends that in \cite{GoldsteinSela} by determining also the $U(1)$ charge of the charged twist operators.

	Considering a single interval $\mathcal{A}$ on the Riemann surface $\mathcal{R}_1$, the partition function $\mathcal{Z}_{n}=\Tr \rho_{\mathcal{A}}^n$ is calculated by evaluating the path integral on the Riemann surface $\mathcal{R}_{n}$, where $\mathcal{R}_n$ is defined as the $n$-sheeted branched covering of $\mathcal{R}_{1}$, with the interval $\mathcal{A}$ as the branch cut on $\mathcal{R}_1$. (Anti-)Twist fields $(\Tilde{\sigma})\ \sigma_n$ are Virasoro primaries of conformal dimension	
	\begin{eqnarray}
		\Delta_n=\bar{\Delta}_n=\frac{c}{24}(n-\frac{1}{n})\ .
	\end{eqnarray}
	defined so as to relate expectation values on $\cR_n$ and $\cR_1$ via
	\begin{eqnarray}
		\langle O(z_1, \bar{z}_1;\text{sheet}\ i)\cdots\rangle_{n}=\langle O_i(z_1,\bar{z}_1)\sigma_n(v_1, \bar{v}_1)\Tilde{\sigma}_n(v_2, \bar{v}_2)\cdots\rangle_{1}\ .\label{relation bt two picture}
	\end{eqnarray} 
	The index $1$ and $n$ denotes the surface $\mathcal{R}_1$ and $\mathcal{R}_n$. Therefore twist fields implement the complicated topology of $\cR_n$ on the single sheet $\cR_1$. The local operator $O(z_1, \bar{z}_1;\text{sheet}\ i)$ lives on the $i^{\text{th}}$-sheet of $\mathcal{R}_n$, which corresponds to the $i^{\text{th}}$-copy of the fields, i.e. $O_i(z_1,\bar{z}_1)$.
	
	Since the non-local operator $e^{i\mu \hat{Q}_\mathcal{A}}$ is equivalent to a pair of vertex operators inserted at the branch points, the charged moments for the vacuum state can be written as
	\begin{eqnarray}
		\mathcal{Z}_{n}(\mu)=\langle V_{\mu}(v_1, \bar{v}_1)V_{-\mu}(v_2, \bar{v_2})\rangle_{n}=\langle \sigma_{n,\mu}(v_1, \bar{v}_1)\Tilde{\sigma}_{n,-\mu}(v_2, \bar{v}_2)\rangle_{1}\ ,\label{vacuum Zn}
	\end{eqnarray}
	The operator $\sigma_{n,\mu}$ and $\Tilde{\sigma}_{n,-\mu}$ on the right hand side of \eqref{vacuum Zn} are the charged twist and anti-charged twist operator, which were first introduced in \cite{Belin:2013uta}. In presence of the pair of vertex operators, the generalization of \eqref{relation bt two picture},can be expressed as
	\begin{eqnarray}
		\langle O(z_1, \bar{z}_1;\text{sheet}\ i) V_{\mu}(v_1, \bar{v}_1)V_{-\mu}(v_2, \bar{v_2})\rangle_{n}=\langle O_i(z_1,\bar{z}_1)\sigma_{n,\mu}(v_1, \bar{v}_1)\Tilde{\sigma}_{n,-\mu}(v_2, \bar{v}_2)\rangle_{1}\ .\label{relation bt two picture 2}
	\end{eqnarray}
	Following the method in \cite{GoldsteinSela}, one finds that the charged twist operator is both a Virasoro primary and current primary. The conformal weight and the $U(1)$ charge of the charged twist operators is given by (cf. \Appref{appendix B})
	\begin{eqnarray}\label{WeightsChargedTwist}
		\Delta_{n,\mu}=\Delta_{n}+\frac{\Delta_{\mu}}{n}\ ,\quad \charge_{\mu,n}=\charge_\mu\ .
	\end{eqnarray}
	The same results hold for the anti-holomorphic part.	
	Here, we clarify the notion of ``fusion"  used in \cite{GoldsteinSela} of an ordinary twist field and a vertex operator to yield a charged twist field. Notice that the conformal weight in \eqref{WeightsChargedTwist} of the vertex operator enters rescaled by $n$, while the charge is not. 
	This suggests that the charged twist operator can be understood as the normal ordered product
	\begin{equation}
		\sigma_{\mu,n}=\normord{\sigma_n V' }, 
		\quad
		\text{with}
		\quad
		V'=\prod_{i=1}^n V^{i}_{\mu/n}\,,
	\end{equation}
	where $V^i_{\mu/n}$ is a vertex operator confined to the $\ith$ copy. From \eqref{confWeightCharge}, it is easy to see that the conformal weight of $V'$ is indeed
	\begin{equation}
		\Delta'=\sum_{i=1}^n\Delta_{\mu/n}^i=\frac{\Delta_\mu}{n},
		\qquad
		\charge'=\sum_{i=1}^n\charge_{\mu/n}^i=\charge_\mu
	\end{equation}
	The product structure of $V'$ implies that the vertex operator is distributed evenly among the sheets of the Riemannian surface $\cR_n$. This clarifies the notion of  ``fusion" used in \cite{GoldsteinSela}.
	
	Given the charged twist operators, the charged moments for the vacuum state are expressed by
	\begin{align}
		\mathcal{Z}_{n}(\mu)&=\left\langle \sigma_{n,\mu}(v_1, \bar{v}_1)\Tilde{\sigma}_{n,-\mu}(v_2, \bar{v}_2) \right\rangle_{1}=\left|\frac{v_1-v_2}{\delta}\right|^{-2(\Delta_{n,\mu}+\bar{\Delta}_{n,\mu})}\nonumber\\
		&=\left|\frac{v_1-v_2}{\delta}\right|^{-\frac{c}{6}(n-\frac{1}{n})-\frac{k}{n}(\frac{\mu}{2\pi})^2}\,.
	\end{align}
	The holomorphic subregion charge can be obtained by integrating the current in \eqref{J n} over the branch cut $\mathcal{A}$, given by
	\begin{align}
		\langle \hat{q}_{\mathcal{A}}\rangle_{n,\mu}&=\int_{v_2+\epsilon}^{v_1-\epsilon}\frac{dz}{2\pi i}\langle \hat{J}(z)\rangle_{n,\mu}=\frac{i\charge_\mu}{n\pi}\ln{\left(\frac{v_1-v_2}{\delta}\right)}\nonumber\\
		&=\frac{i k\mu}{4\pi^2 n}\ln{\left(\frac{v_1-v_2}{\delta}\right)}\ ,
	\end{align}
	and the total subregion charge reads
	\begin{eqnarray}
		\langle \hat{Q}_{\mathcal{A}}\rangle_{n,\mu}=\frac{i k\mu}{2\pi^2 n}\ln{\left|\frac{v_1-v_2}{\delta}\right|}\ .
	\end{eqnarray}
	The resulting symmetry-resolved R\'enyi entropy is given by
	\begin{align}
		S_{n}(q)=\frac{c}{6}(1+\frac{1}{n})\ln{\left|\frac{v_1-v_2}{\delta}\right|}-\frac{1}{2}\ln{\left(\frac{k}{\pi} \ln{\left|\frac{v_1-v_2}{\delta}\right|}\right)}+O(1)\ ,
	\end{align}
	and the symmetry-resolved entanglement entropy reads
	\begin{align}\label{see vacuum}
		S(q)=\frac{c}{3}\ln{\left|\frac{v_1-v_2}{\delta}\right|}-\frac{1}{2}\ln{\left(\frac{k}{\pi} \ln{\left|\frac{v_1-v_2}{\delta}\right|}\right)}+O(1)\ .
	\end{align}
	This confirms our holographic result \eqref{chargedAdSSREE}. This equipartition of the entanglement entropy is in fact universal even in excited states \cite{Capizzi:2020jed}, which can be seen as follows. The singular terms in the OPEs of $J$ with any current primaries contain only simple poles, with their coefficients given by the charges of the primary fields. Since the subregion charge is obtained by taking the integration of the current over the branch cut, the divergent part of the subregion charge is always determined by the charged twist operators, located at the branch points. As a result, these lead to the universal behavior \eqref{see vacuum} of the symmetry-resolved entanglement entropy.
	
	\section{Discussion and Outlook}\label{Section 6}
	In this paper we extended the  framework for computing the symmetry-resolved entanglement entropy by introducing a new generating function method. It allows to relate the charge part of the symmetry-resolved entanglement entropy to the expectation value of the subregion charge operator $\langle Q_\cA\rangle$. Our method is particularly useful in the holographic context, since it  reduces the computation of the symmetry resolved entropies to the calculation of the boundary subregion charge $\langle Q_\cA\rangle $. As we show explicitly for the case of $U(1)$ Chern-Simons theory coupled to three-dimensional Einstein-Hilbert gravity with negative cosmological constant, this simplification in particular enables us to derive the symmetry-resolved entanglement entropy without the necessity of deriving the full renormalization of the holographic dual. 
	
	In the context of $U(1)$ Chern-Simons gravity, our holographic construction is inspired by the field theory picture of \cite{GoldsteinSela}, in which an Aharanov-Bohm flux is introduced on the replica manifold. We realized this flux in asymptotically $\ads$ geometries by inserting its natural holographic dual, namely a charged Wilson line defect \eqref{gravityDefect} at the fixed point locus of the replica symmetry. This defect sources the $U(1)$ Chern-Simons gauge field, from which we calculate the subregion charge $\langle Q_\cA\rangle$. In turn, the subregion charge allowed us to determine the generating function  \eqref{generatingfunction}. This result was used to directly calculate the interesting charge contribution to the symmetry resolved entanglement entropy. The charge-dependent partition function of the dual CFT differs from the generating function by the insertion of an operator, c.f.~\eqref{pt} and \eqref{Delta0}, which however cancels out in the calculation of the subregion charge, c.f.~\eqref{Q-eq}. This cancellation relies on the structure of the $\u{(1)_k}$ Kac-Moody algebra, which implies that the additional operator indeed commutes with the subregion charge. As a consequence, the subregion charge $\langle Q_\cA\rangle$ can be obtained independent of knowing the charged moments. The subregion charge together with the Ryu-Takayanagi formula \cite{RT} for entanglement entropy can in turn be used to determine the full symmetry-resolved entanglement entropy without the necessity to directly calculate the charged moments. 
	
	Besides the calculation of the symmetry resolved entanglement entropy in Poincar\'e AdS, we also applied our generating function method to general asymptotic $\ads_3$ backgrounds. We presented further explicit calculations of the symmetry-resolved entanglement entropy for the examples of global $\ads_3$, and conical defect space-times. Our results show that in all these cases, the probability distributions \eqref{Pn} for the subregion charge fluctuations are always Gaussian. Furthermore, the resulting symmetry-resolved entanglement entropies exhibit a universal equipartition behavior, i.e. the entanglement for each charge sector is identical and depends only on the geodesic lengths given by the Ryu-Takayanagi prescription \cite{RT} and the Chern-Simons level $k$. For more complicated symmetry algebrae, we suspect a breakdown of the equipartition of entanglement.

	Since the asymptotic symmetry of the $U(1)$ Chern-Simons theory is of $\u{(1)_k}$ Kac-Moody type, in our CFT calculations we assume the dual CFT to admit the same symmetry algebra. Using this summetry, we show that the non-local operator \eqref{hol non-local I} can be identified with a pair of charged $U(1)$ vertex operators, up to a cut-off dependent proportionality constant. As expected from \cite{GaberdielKlemmRunkel},  this pair of vertex operators generates $U(1)$ charge defects on the multi-sheeted Riemann surface. Working in the twist operator picture, we use the product of the charged vertex operator with the usual twist operator $\sigma_n$ to define a charged twist operator $\sigma_{n,\mu}$. Following the method in \cite{GoldsteinSela}, we find that this charged twist operator is both a Virasoro primary and a current primary. The charged twist operator allows us to compute the charged moments and the expectation value of the subregion charge for the vacuum state in a two-dimensional CFT with $\u{(1)_k}$ Kac-Moody symmetry. Our CFT calculations are in complete agreement with the holographic results.

	Our work, and in particular the generating function method, allows for several natural extensions. One option is to investigate the case of $m$ intervals, in particular also in view of possible topological transitions in the symmetry resolved entanglement and R\'enyi entropies, along the lines of \cite{RangamaniTakayanagiBook,Abt:2017pmf}. The case of $m$ intervals reduces to the calculation of CFT correlators with $2m$ insertions of charged twist operators. It will in particular be interesting to see whether the subregion charges in different intervals can be chosen independently, or are correlated with each other in some way. Another interesting quantum information measures that could be resolved in different charge sectors is the mutual information, which should extract the non-divergent piece of the symmetry-resolved entanglement entropy. Another finite entanglement measure is the relative entropy \cite{MyersCasiniBlanco}. 
	Other interesting extensions of this work involve non-abelian Aharanov-Bohm fluxes. These should be captured by defects of non-abelian Wess-Zumino-Witten type \cite{Bachas:2004sy,Erdmenger:2020hug}. In the context of holography, an investigation of higher spin gravity will be interesting as well \cite{toappear}. Since our bulk theory admits the $\u{(1)_k}$ Kac-Moody symmetry which also governs the CFT description of abelian anyons, it would furthermore be interesting to compare our result to the symmetry resolved entanglement in anyonic conformal field theories. 
	In all these cases, we expect our generating function method to considerably simplify the calculation of symmetry resolved entanglement measures. Finally, three-dimensional gravity coupled $U(1)$ Chern-Simons gauge fields have emerged in the study of two-dimensional free conformal field theories over the Narain lattice \cite{Afkhami-Jeddi:2020ezh,Maloney:2020nni,Perez:2020klz,Dymarsky:2020pzc}. Studying the symmetry-resolved entanglement entropy in this setting may lead to a refined insight into this novel instance of a holographic duality. Relatedly, developing a proof of our prescription for the holographic dual \eqref{gravityDefect} of the charged moments along the lines of \cite{Lewkowycz:2013nqa, Dong:2016fnf} would further add to the understanding of the holographic dual of the symmetry-resolved entanglement entropy.

	\acknowledgments
	We thank Martin Ammon, Souvik Banerjee, Pascal Fries and Konstantin Weisenberger for useful discussions. 
	R.M., C.N. and S.Z. acknowledge support by the Deutsche Forschungsgemeinschaft (DFG, German Research Foundation) under Germany's Excellence Strategy through the W\"urzburg‐Dresden Cluster of Excellence on Complexity and Topology in Quantum Matter ‐ ct.qmat (EXC 2147, project‐id 390858490). The work of R.M. and C.N. was furthermore supported via project id 258499086 - SFB 1170 ’ToCoTronics’. S.Z. is financially supported  by the China Scholarship Council.
	

	\begin{appendix}
		\section{Asymptotic symmetry algebra}\label{appendix A}
		The asymptotic symmetry algebra can be obtained by the standard Hamiltonian approach \cite{Regge:1974zd, Banados:1994tn} or the covariant phase space approach \cite{Crnkovic:1986ex, Lee:1990nz, Wald:1993nt, Harlow:2019yfa}. Here we work in the latter framework. Consider a hyper-surface $\Sigma$, which is defined as the base manifold of the unconstrained phase space. The variation of the pure Chern-Simons action gives
		\begin{eqnarray}
			\delta I_{CS}=\frac{ik}{4\pi}\int \delta A\wedge dA+\frac{ik}{8\pi}\int_{\Sigma} \delta A\wedge A\label{var CS}
		\end{eqnarray}
		Here, $\delta$ represents the exterior derivative in the phase space and $\delta A$ is a one-form on phase space. A second variation of the boundary term in \eqref{var CS} leads to a closed and non-degenerate symplectic two-form on the unconstrained phase space, given by
		\begin{eqnarray}
			\omega=\frac{ik}{8\pi}\int_{\Sigma}\delta A\wedge\delta A\,.
		\end{eqnarray}
		In components this reads
		\begin{eqnarray}
			\omega=\frac{1}{2}\omega^{i j}\delta A_i\wedge\delta A_j\,.
		\end{eqnarray}
		Then the Poisson bracket of two functionals $F[A],\,H[A]$ on the unconstrained phase space is defined by
		\begin{eqnarray}
			\{F,H\}&:=&i\omega_{i j}\frac{\partial F}{\partial A_i}\frac{\partial H}{\partial A_j} =\frac{4\pi}{ k}\int_{\Sigma}\Tilde{\epsilon}_{i j}\frac{\partial F}{\partial A_i}\frac{\partial H}{\partial A_j}\label{Poisson bracket def}\,,
		\end{eqnarray}
		where $\omega_{ij}$ is the inverse map of $\omega^{i j}$ and $\Tilde{\epsilon}_{ij}$ is defined on $\Sigma$ with chosen orientation such that $\Tilde{\epsilon}_{12}=1$ in the local coordinates.
		There is a so-called smeared function $G(\eta)$ on the unconstrained phase space, which is associated with a infinitesimal gauge parameter $\eta$, and generates the gauge transformation $A\to A+d\eta$, given by
		\begin{eqnarray}
			G(\eta)=\frac{ k}{4\pi}\int_{\Sigma}A\wedge d\eta\,.
		\end{eqnarray}
		Indeed, applying the definition \eqref{Poisson bracket def}, we find
		\begin{eqnarray}
			\{G(\eta), A\}=d\eta\,.
		\end{eqnarray}
		Computing the Poisson bracket of two smeared functions associated with two gauge parameters, we find
		\begin{eqnarray}
			\{G(\eta),G(\lambda)\}=\frac{k}{4\pi}\int_{\Sigma}d\eta\wedge d\lambda.\label{G poisson}
		\end{eqnarray}
		The smeared function can be decomposed into
		\begin{eqnarray}
			G(\eta)= Q(\eta)+C(\eta)
		\end{eqnarray}
		with
		\begin{eqnarray}
			Q(\lambda)=-\frac{ k}{4\pi}\int_{\partial\Sigma}\eta A, \quad C(\lambda)=\frac{k}{4\pi}\int_{\Sigma}\eta dA\,.
		\end{eqnarray}
		After imposing the constraints $C=0$, the charge $Q$ becomes the generator of the gauge transformation on the constrained phase space of on-shell solutions, and the Poisson bracket \eqref{G poisson} reduces to the Poisson bracket of the charge $Q$,
		\begin{eqnarray}
			\{Q(\eta),Q(\lambda)\}=\frac{k}{4\pi}\int_{\partial\Sigma}\eta d\lambda\,.
		\end{eqnarray}
		In order to be compatible with the variational principle \eqref{var A} as well as the boundary condition \eqref{A expansion}, the gauge parameters are required to have the following asymptotic behavior
		\begin{eqnarray}
			\eta=\eta^{(0)}(w)+e^{-2\rho}\eta^{(2)}(\rho,w,\bar{w})+\cdots, \quad \rho\to\infty\,.
		\end{eqnarray}
		Therefore, if we choose $\Sigma$ to be a constant time slice, we obtain the associated charge
		\begin{eqnarray}
			Q(\eta)=\frac{ k}{4\pi}\oint d\phi\ \eta^{(0)} \left(A^{(0)}_{w}+A^{(0)}_{\bar{w}}\right).
		\end{eqnarray}
		It follows from the variational principle that $A_{\bar{w}}$ fixed, which means it is a Lie algebra constant. Hence, if we choose a gauge fixing such that $A^{(0)}_{w}$ $(A^{(0)}_{\bar{w}})$ are holomorphic (anti-holomorphic), then the Poisson bracket is further reduced to
		\begin{eqnarray}
			\{\appq(\eta),\appq(\lambda)\}=\frac{k}{4\pi}\oint dw \ \eta^{(0)} \partial_w\lambda^{(0)},\label{q poisson}
		\end{eqnarray}
		with the holomorphic charge $q(\eta)$ defined as
		\begin{eqnarray}
			\appq(\eta)=\frac{k}{4\pi}\oint dw \ \eta^{(0)}A^{(0)}_{w}=\frac{1}{2\pi i}\int dw \ \eta^{(0)}J(w)\label{holo charge}\,.
		\end{eqnarray}
		Here we denote $J_w=J(w)$ since it is holomorphic by gauge fixing. This holomorphic charge is the generator of the residual gauge transformation arising from the gauge fixing.
		
		The gauge parameters have a mode expansion
		\begin{eqnarray}
			\eta^{(0)}=\sum_{n=-\infty}^{\infty}\eta_n e^{-in w},\quad \lambda^{(0)}=\sum_{n=-\infty}^{\infty}\lambda_n e^{-in w}.\label{gauge modes}
		\end{eqnarray}
		Combining with the mode expansion in \eqref{current modes} and inserting them into \eqref{holo charge}, we obtain a simple form for the charge
		\begin{eqnarray}
			\appq(\eta)=\frac{1}{2\pi i}\oint dw\ \eta^{(0)} J(w)=-i\sum_{n=-\infty}^{\infty}\eta_n J_n\label{Q1}\,.
		\end{eqnarray}
		Inserting \eqref{Q1} into \eqref{q poisson}, and evaluating the right hand side, we find
		\begin{eqnarray}
			i\{J_n, J_m\}=\frac{k}{2}n\delta_{n+m,0}\,,
		\end{eqnarray}
		which is the ``classical" $\u(1)_{k}$ affine Lie algebra. The quantum version of this algebra is obtained by replacing $i\{,\}$ by commutators $[,]$, yielding the $\u(1)_{k}$ Kac-Moody algebra
		\begin{eqnarray}
			[J_n, J_m]=\frac{k}{2}n\delta_{n+m,0}.
		\end{eqnarray}
		
		Next, we verify that the modified stress tensor $T_{ww}[J]$ \eqref{TJ} indeed satisfies the Virasoro algebra. In particular, we show that it will not shift the central charge on the gravity side. This is expected since the $\u(1)_k$ model has central charge one and therefore does not alter the large central charge of the gravity sector; see below.
		
		The stress tensor $T_{ww}[J]$ is the generator of infinitesimal asymptotic diffeomorphisms \eqref{asy diffeo} for the gauge field, which is actually equivalent to a corresponding field-dependent gauge transformation when $dA^{(0)}=0$. This fact can be easily shown by the Cartan formula,
		\begin{eqnarray}
			\delta_{\xi}A=\mathcal{L}_{\xi}A=d(\iota_{\xi}A)+\iota_{\xi}dA\approx d \alpha +\iota_{\xi}dA^{(0)},  \quad \rho\to\infty,
		\end{eqnarray}
		with the corresponding gauge parameter
		\begin{eqnarray}
			\alpha=\xi(w)A^{(0)}(w)+\bar{\xi}(\bar{w})A^{(0)}_{\bar{w}},
		\end{eqnarray}
		Note that in this case, the requirement of holomorphic gauge parameter $\alpha$ enforces the vanishing of the source $A^{(0)}_{\bar{w}}$, which is similar to the situation in the gravity sector, where the symmetry algebra is broken if the gravitational source appears.
		
		If we impose $A^{(0)}_{\bar{w}}=0$, then in this case, the corresponding diffeomorphism charge $\diffq$ associated with the vector field $\xi^{\mu}$ is given by
		\begin{eqnarray}
			\diffq(\alpha)=\frac{ k}{8\pi}\oint dw \ \xi(w)A^{(0)}_{w} A^{(0)}_{w}=\oint dw \ \xi(w)T_{ww}[J]\ .\label{q alpha}
		\end{eqnarray}
		Expanding the $T_{ww}[J]$ and $\xi(w)$ in modes, defined through
		\begin{eqnarray}
			L'_n=-\oint dw \ e^{-inw}T_{ww}[J],\quad \xi(w)=\sum_{n=-\infty}^{\infty}\xi_n e^{-inw}\,,
		\end{eqnarray}
		the simplified form of the diffeomorphism charge is obtained as
		\begin{eqnarray}
			\diffq(\alpha)=-\sum_{n=-\infty}^{\infty} \xi_n L_n.\label{q alpha modes}
		\end{eqnarray}
		With this, the Poisson bracket \eqref{q poisson} of two diffeomorphism charges $\diffq(\alpha)$ and $\diffq(\beta)$, reduces to the Witt algebra
		\begin{eqnarray}
			i\{L'_n, L'_m\}=(n-m)L'_{n+m}\ .\label{Witt}
		\end{eqnarray}
		Similarly, inserting \eqref{Q1} and \eqref{q alpha} into \eqref{q poisson}, the Poisson bracket of diffeomorphism charge $q(\alpha)$ and charge $q(\eta)$ yields
		\begin{eqnarray}
			i\{L'_n, J_m\}=-m J_{n+m}\ .
		\end{eqnarray}
		Combining the above algebrae with the gravitational Virasoro algebra \eqref{cl vir}, one obtains the full asymptotic symmetry algebra of the theory,
		\begin{eqnarray}
			i\{J_{n}, J_m\}&=&\frac{1}{2}n k \delta_{m+n,0},\nonumber\\
			i\{L_{n}, J_m\}&=& -m J_{n+m},\nonumber\\
			i\{L_{n}, L_m\}&=&(n-m)L_{n+m}+\frac{c}{12}(n^3-n)\delta_{n+m,0}.
		\end{eqnarray}
		and analogously for the anti-holomorphic part.
		
		The quantum version of this asymptotic symmetry algebra is obtained after replacing $i\{,\}$ by the commutators $[,]$, and should be identified as the symmetry algebra in the dual CFT by the AdS/CFT correspondence. It is interesting to note that the effective central charge in the dual CFT does not shift even in presence of the $U(1)$ Chern-Simons fields on the gravity side. This fact essentially arises from the boundary condition, i.e. $A^{(0)}_{\rho}=0$, which yields no central term contributions to the diffeomorphism charge in \eqref{q alpha}. From the perspective of the AdS/CFT correspondence, this is also expected, since it is well-known that the CFT stress tensor constructed from a $U(1)$ Kac-Moody current by the Sugawara construction satisfies the OPE \cite{Goddard:1986bp}
		\begin{eqnarray}
			\frac{1}{k}(\hat{J}\hat{J})(z)\frac{1}{k}(\hat{J}\hat{J})(w)\sim\frac{1/2}{(z-w)^4}+\cdots\ ,
		\end{eqnarray}
		revealing a central charge contribution $c_J=1$ to the full theory. Hence it can be neglected in large $c$ limit.
		
		\section{Vertex operators and charged twist operators}\label{appendix B}
		\subsection*{Non-local operator}
		The radial ordering of the left-moving non-local operator $I_{\mu}(v_1,v_2)$, defined in \eqref{hol non-local I} can be written as
		\begin{eqnarray}
			\hat{R}[I_{\mu}(v_1,v_2)]
			=
			\sum_{n=0}^{\infty}\frac{1}{n!}(\frac{\mu}{2\pi})^{n}\int_{v_2}^{v_1}\cdots\int_{v_2}^{v_1}dz_1\cdots dz_n \hat{R}\left[J(z_1)\cdots J(z_n)\right]\ ,\label{I 1}
		\end{eqnarray}
		$\hat{R}$ is the radial ordering operator. Wick's theorem implies the relation
		\begin{eqnarray}
			\hat{R}\left[J(z)J(w)\right]=\normord{J(z)J(w)}+\frac{k/2}{(z-w)^2}\ .
		\end{eqnarray}
		Therefore we have
		\begin{align}
			\int_{v_2}^{v_1}\int_{v_2}^{v_1}dz dw\hat{R}\left[J(z)J(w)\right]&=\int_{v_2}^{v_1}\int_{v_2}^{v_1}dz dw\left(\normord{J(z)J(w)}+\frac{k/2}{(z-w)^2}\right)\ ,\nonumber\\
			&=\int_{v_2}^{v_1}\int_{v_2}^{v_1}dz dw\normord{J(z)J(w)}\nonumber\\
			&\quad +\int_{v_2+\delta}^{v_1-\delta}dz\left(\frac{k/2}{z-v_1}-\frac{k/2}{z-v_2}\right)\nonumber\\
			&\approx \int_{v_2}^{v_1}\int_{v_2}^{v_1}dz dw\normord{J(z)J(w)}-k\ln{\left(\frac{v_1-v_2}{\delta}\right)}\label{radial ordering relation 1}\,.
		\end{align}
		For simplicity, we denote
		\begin{align}
			X_n&=\int_{v_2}^{v_1}\cdots\int_{v_2}^{v_1}dz_1\cdots dz_n \hat{R}\left[J(z_1)\cdots J(z_n)\right],\\
			Y_n&=\int_{v_2}^{v_1}\cdots\int_{v_2}^{v_1}dz_1\cdots dz_n \normord{J(z_1)\cdots J(z_n)}.
		\end{align}
		and
		\begin{eqnarray}
			B=-k\ln{\left(\frac{v_1-v_2}{\delta}\right)}
		\end{eqnarray}
		Then we can write down the generalization of \eqref{radial ordering relation 1} as
		\begin{eqnarray}
			X_n=N_{n,0}Y_{n}+N_{n,1} Y_{n-2}B+N_{n,2} Y_{n-4}B^2+...+N_{n,m}Y_{n-2m}B^{m}+\cdots
		\end{eqnarray}
		where $N_{n,m}$ is the number of terms for $m$ times contractions of $X_n$, given by
		\begin{align}
			N_{n,m}&=\frac{1}{m!}C_{n}^{2m}\cdot C_{2m}^{2m-2}\cdot C_{2m-2}^{2m-4}\cdots C_{4}^{2}=\frac{n!}{2^{m}m!(n-2m)!}\ .
		\end{align}
		By summing over all the terms with the same number of contractions in \eqref{I 1}, we can rewrite the radially ordered non-local operator as
		\begin{align}
			R[I_{\mu}(v_1,v_2)]&=\left(\sum_{n=0}^{\infty}\frac{1}{n!}(\frac{\mu}{2\pi})^{n}Y_{n}\right)\cdot\left(\sum_{m=0}^{\infty}(\frac{\mu}{2\pi})^{2m}(\frac{B}{2})^{m}\right)\nonumber\\
			&=\left(\frac{v_1-v_2}{\delta}\right)^{-\frac{k}{2}(\frac{\mu}{2\pi})^{2}}\normord{I_{\mu}(v_1, v_2)}
		\end{align}
		
		\subsection*{Charge and conformal weight of vertex operator}
		In this subsection we derive the OPEs of the current and stress tensor with charged $U(1)$ vertex operators. We do so without the Sugawara construction since we do not assume knowledge of the exact CFT dual.
		
		To get the OPEs for the vertex operators, we integrate the OPEs \eqref{OPE1}, leading to
		\begin{eqnarray}
			\hat{T}(z)\chi(w)& \sim & \frac{\hat{J}(w)}{z-w}+\cdots\ ,\nonumber\\
			\hat{J}(z)\chi(w)& \sim & \frac{k/2}{z-w}+\cdots\ .
		\end{eqnarray}
		Then, using the generalized Wick Theorem, we obtain the OPEs for the vertex operator
		\begin{eqnarray}
			\hat{J}(z)V_{\mu}(0)&\sim &\frac{k}{2}\left(\frac{\mu}{2\pi}\right)\frac{V_{\mu}(0)}{z}+\cdots\nonumber\\
			\hat{T}(z)V_{\mu}(0)&=&\sum_{n=0}^{\infty}\left[\frac{(\mu/2\pi)^n}{n!}\left(\prod_{i=1}^{n-1}\oint_{w_i=0} \frac{d w_{i}}{2\pi i w_i}\right) \hat{T}(z)\chi(w_1)\chi(w_2)\cdots\chi(w_{n-1}) \chi(0)\right]+\cdots\nonumber\\
			&\sim & \sum_{n=0}^{\infty}\left[\frac{(\mu/2\pi)^n}{(n-1)!}\left(\prod_{i=1}^{n-1}\oint_{w_i=0} \frac{d w_{i}}{2\pi i w_i}\right) \frac{\hat{J}(w_1)}{z-w_1}\chi(w_2)\cdots \chi(0)\right]+\cdots\nonumber\\
			&\sim & \sum_{n=0}^{\infty}\left[\frac{(\mu/2\pi)^n}{(n-1)!}\left(\prod_{i=1}^{n-1}\oint_{w_i=0} \frac{d w_{i}}{2\pi i w_i}\right) \left(\frac{\hat{J}(w_1)}{z}+\frac{w_1 \hat{J}(w_1)}{z^2}+\cdots\right)\chi(w_2)\cdots \chi(0)\right]+\cdots\nonumber\\
			&\sim & \sum_{n=0}^{\infty}\left[\frac{(\mu/2\pi)^n}{(n-1)!}\left(\prod_{i=1}^{n-1}\oint_{w_i=0} \frac{d w_{i}}{2\pi i w_i}\right) \frac{\hat{J}(w_1)}{z}\chi(w_2)\cdots \chi(0)\right]\nonumber\\
			&+&\sum_{n=0}^{\infty}\left[\frac{(\mu/2\pi)^n}{(n-2)!}\cdot\frac{1}{2 z^2}\left(\prod_{i=1}^{n-1}\oint_{w_i=0} \frac{d w_{i}}{2\pi i w_i}\right) \left(\frac{k w_1}{2(w_1-w_2)}-\frac{k w_2}{2(w_1-w_2)}\right)\chi(w_3)\cdots \chi(0)\right]+\cdots\nonumber\\
			&\sim & \frac{k}{4}\left(\frac{\mu}{2\pi}\right)^2\frac{V_{\mu}(0)}{z^2}+\frac{\partial V_{\mu}(0)}{z}+\cdots\ .
		\end{eqnarray}
		
		\subsection*{Charged twist operator}
		This section presents a thorough derivation of the conformal weight and the $U(1)$ charge of the charged twist operators $\sigma_{n,\mu}$ using the original $n$-sheeted geometry $\cR_n$. The conformal weight is already given in \cite{GoldsteinSela}, while the charge is calculated explicitely here for the first time.

		When $\mathcal{R}_1$ is a flat complex plane with complex coordinates $z$ and $\bar{z}$, the $n$-sheeted branched covering $\mathcal{R}_{n}$ can be mapped to a flat $z'$-complex plane $C$ by the uniformization map \cite{Calabrese:2004eu},
		\begin{eqnarray}\label{uniformizationMap}
			z'=\left(\frac{z-v_2}{z-v_1}\right)^{\frac{1}{n}}\ .
		\end{eqnarray}
		The expectation values for the stress tensor and the current on $C$ are given by
		\begin{align}
			\langle \hat{T}(z') \rangle_{C,\mu}
			=
			\lim_{w\to\infty}\frac{\langle \hat{T}(z') V_{\mu}(w)V_{-\mu}(0) \rangle_{C}}{\langle V_{\mu}(w)V_{-\mu}(0) \rangle_{C}}
			=
			\lim_{w\to\infty}\Delta_{\mu}\frac{(w-0)^2}{(z'-w)^2(z'-0)^2}
			=
			\frac{\Delta_{\mu}}{z'^2}\ ,
		\end{align}
		and
		\begin{eqnarray}
			\langle \hat{J}(z') \rangle_{C,\mu}
			=
			\lim_{w\to\infty}\frac{\langle \hat{J}(z') V_{\mu}(w)V_{-\mu}(0) \rangle_{C}}{\langle V_{\mu}(w)V_{-\mu}(0) \rangle_{C}}
			=
			\lim_{w\to\infty}\frac{\charge_\mu}{z'-w}+\frac{\charge_{-\mu}}{z'-0}
			=
			-\frac{\charge_\mu}{z'}\, .
		\end{eqnarray}
		Then, by the transformation laws of the stress tensor and current under the conformal transformation \eqref{uniformizationMap}, one obtains the corresponding expectation values on $\mathcal{R}_n$, given by
		\begin{eqnarray}
			\langle \hat{T}(z) \rangle_{n,\mu}=\left(\frac{dz'}{dz}\right)^{2}\langle \hat{T}(z') \rangle_{C,\mu}+\frac{c}{12}\{z',z\}=\left(\frac{\Delta_n}{n}+\frac{\Delta_{\mu}}{n^2}\right)\frac{(v_1-v_2)^2}{(z-v_1)^2(z-v_2)^{2}}\ ,\label{T n}
		\end{eqnarray}
		and
		\begin{eqnarray}
			\langle \hat{J}(z) \rangle_{n,\mu}=\left(\frac{dz'}{dz}\right)\langle \hat{J}(z') \rangle_{C,\mu}=\frac{\alpha_\mu}{n}\left(\frac{1}{z-v_1}-\frac{1}{z-v_2}\right)\ .\label{J n}
		\end{eqnarray}
		Coming back to the twist field picture, \eqref{relation bt two picture 2} yields
		\begin{eqnarray}
			\langle \hat{T}(z) \rangle_{n,\mu}=\frac{\langle \hat{T}_i(z) \sigma_{n,\mu}(v_1, \bar{v}_1)\Tilde{\sigma}_{n,-\mu}(v_2, \bar{v}_2)\rangle_{1}}{\langle \sigma_{n,\mu}(v_1, \bar{v}_1)\Tilde{\sigma}_{n,-\mu}(v_2, \bar{v}_2) \rangle_{1}}=\langle \hat{T}_{i}(z) \rangle_{1,\mu} \ ,
		\end{eqnarray}
		and
		\begin{eqnarray}
			\langle \hat{J}(z) \rangle_{n,\mu}=\frac{\langle \hat{J}_i(z) \sigma_{n,\mu}(v_1, \bar{v}_1)\Tilde{\sigma}_{n,-\mu}(v_2, \bar{v}_2)\rangle_{1}}{\langle \sigma_{n,\mu}(v_1, \bar{v}_1)\Tilde{\sigma}_{n,-\mu}(v_2, \bar{v}_2) \rangle_{1}}=\langle \hat{J}_{i}(z) \rangle_{1,\mu}\ ,
		\end{eqnarray}
		for $i=1,\dots,n$, which gives the stress tensor and current for a single copy of the fields. Multiplying \eqref{T n} and \eqref{J n} by $n$ provides the total stress tensor and current in the n copies of CFT, given by
		\begin{eqnarray}
			\langle \hat{T}^{(n)}(z) \rangle_{1,\mu} = \sum_{i=1}^{n}\langle \hat{T}_{i}(z) \rangle_{1,\mu} =\left(\Delta_n+\frac{\Delta_{\mu}}{n}\right)\frac{(v_1-v_2)^2}{(z-v_1)^2(z-v_2)^{2}}\ ,
		\end{eqnarray}
		and
		\begin{eqnarray}
			\langle \hat{J}^{(n)}(z) \rangle_{1,\mu} = \sum_{i=1}^{n}\langle \hat{J}_{i}(z) \rangle_{1,\mu} =\frac{\charge_\mu}{z-v_{1}}-\frac{\charge_\mu}{z-v_2}\ .
		\end{eqnarray}
		Comparing with the Ward identities of the conformal and $U(1)$ symmetries, one can read off the conformal dimension and the charge for the charged twist operator,
		\begin{eqnarray}
			\Delta_{n,\mu}=\Delta_{n}+\frac{\Delta_{\mu}}{n}\ ,\quad \charge_{\mu,n}=\charge_\mu\ .
		\end{eqnarray}
		The same results hold for the anti-holomorphic part.
	\end{appendix}
	\bibliographystyle{JHEP}
	\bibliography{../library}
	
\end{document}